\documentclass[manuscript]{acmart}
\AtBeginDocument{%
  }

\setcopyright{acmlicensed}
\copyrightyear{2026}
\acmYear{2026}
\acmDOI{XXXXXXX.XXXXXXX}

\acmConference[Conference acronym 'XX]{Make sure to enter the correct
  conference title from your rights confirmation email}{June 03--05,
  2018}{Woodstock, NY}
\acmISBN{978-1-4503-XXXX-X/18/06}

\begin{document}

\title{Relying on LLMs: Student Practices and Instructor Norms are Changing in Computer Science Education}
%When Teachers Say "Don't Answer": Negotiating Directness vs. Scaffolding in LLM-Mediated CS Education

\author{Xinrui Lin}
\email{x.lin@bit.edu.cn}
\orcid{0000-0003-3632-5563}
% \authornotemark[3]
\affiliation{%
  \institution{Beijing Institute of Technology}
  \city{Beijing}
  %\state{}
  \country{China}}

\author{Heyan Huang}
\authornote{Corresponding author.}
\email{hhy63@bit.edu.cn}
\orcid{0000-0002-0320-7520}
\affiliation{%
  \institution{Beijing Institute of Technology}
  \city{Beijing}
  %\state{}
  \country{China}}

\author{Shumin Shi}
\email{bjssm@bit.edu.cn}
\orcid{0000-0003-3436-7575}
\affiliation{%
  \institution{Beijing Institute of Technology}
  \city{Beijing}
  %\state{}
  \country{China}}

\author{John Vines}
\authornotemark[1]
\email{john.vines@ed.ac.uk}
\orcid{0000-0003-4051-3356}
\affiliation{%
  \institution{University of Edinburgh}
  \city{Edinburgh}
  %\state{}
  \country{United Kingdom}}

\renewcommand{\shortauthors}{Lin et al.}

\begin{abstract}
%我们先与16个计算机科学学生进行用户研究，收集了他们在不同学习场景下使用LLM的用例，了解了不同场景下他们对待LLM生成内容的实践与态度。我们将这些使用场景划分为写作，解题，信息检索，基于项目的学习，代码（代码题与代码项目），每个场景都归纳了学生的LLM使用用例和他们的见解。接着，我们采访了6个计算机科学老师并邀请他们对学生的实践与态度做出，通过数据分析，我们总结了不同场景下老师对学生依赖LLM的边界设定。最后，针对每个场景，我们讨论了学生行为与老师期望的冲突，老师对LLM影响下教学方式变化的看法，以及如何设计LLM来平衡师生的需求。
Prior research has raised concerns about students' over-reliance on large language models (LLMs) in higher education. This paper examines how Computer Science students and instructors engage with LLMs across five scenarios: "Writing", "Quiz", "Programming", "Project-based learning", and "Information retrieval". Through user studies with 16 students and 6 instructors, we identify 7 key intents, including increasingly complex student practices. Findings reveal varying levels of conflict between student practices and instructor norms, ranging from clear conflict in "Writing-generation" and "(Programming) quiz-solving", through partial conflict in "Programming project-implementation" and "Project-based learning", to broad agreement in "Writing-revision \& ideation", "(Programming) quiz-correction" and "Info-query \& summary". We document instructors are shifting from prohibiting to recognizing students’ use of LLMs for high-quality work, integrating usage records into assessment grading. Finally, we propose LLM design guidelines: deploying default guardrails with game-like and empathetic interaction to prevent students from "deserting" LLMs, especially for "Writing-generation", while utilizing comprehension checks in low-conflict intents to promote learning.

\end{abstract}




\received{12 September 2025}
\received[revised]{10 December 2025}
\received[accepted]{16 January 2026}

\maketitle

\section{Introduction}
%第一段：学生依赖大模型，大模型是双刃剑，可以提高学习效率，使学生仅需很小的努力获得答案
%第二段：研究发现学生依赖大模型获得答案或造成问题，以及研究人如何用交互技术解决这种挑战。
%第三段，引出研究空白，解释为什么研究，对人机交互社区和高等教育社区有什么好处？
%第四段，我们的研究问题

Large language models (LLMs) like ChatGPT are rapidly reshaping learning practices in higher education. Many students now turn to LLMs for assistance with tasks ranging from information retrieval and problem-solving to writing and programming \cite{park_promise_2024, stojanov_university_2024}. Early reports have highlighted the extent of this reliance, indicating that nearly 90\% of students use ChatGPT to complete their assignments \cite{westfall_educators_2023}. Current scholarly discussion often characterizes this phenomenon through a binary lens, viewing LLMs as a double-edged sword \cite{prather_widening_2024, ghimire_generative_2024, mogavi_chatgpt_2024}. 
On one hand, LLMs offer on-demand, highly plausible answers in natural language, thereby accelerating productivity and learning efficiency \cite{mogavi_chatgpt_2024, youssef_examining_2024, park_promise_2024}. Specifically, LLMs are perceived to hold the potential to transform education in \textbf{scenarios} such as writing \cite{lee_coauthor_2022, sultanum_datatales_2023, dhillon_shaping_2024, cotton_chatting_2024}, programming \cite{becker_programming_2023, prather_widening_2024, yan_llm-based_2025}, and project-based learning \cite{zheng_charting_2024, ravi_co-designing_2025}, supported by highly personalized instruction \cite{kasneci_chatgpt_2023, baig_chatgpt_2024, park_promise_2024}. 
On the other hand, these generated essays and answers are often deemed sufficient to pass university assignments \cite{malinka_educational_2023}, allowing students to obtain direct solutions with minimal effort and bypass the deep thinking that traditional problem-solving requires. Evidence suggests that this reliance encourages superficial learning and undermines the acquisition of fundamental skills \cite{kasneci_chatgpt_2023, jost_impact_2024, bastani_generative_2025, rogers_chatgpts_2025}, erodes students' cognitive abilities and critical thinking \cite{mogavi_chatgpt_2024, stadler_cognitive_2024, youssef_examining_2024, lee_impact_2025, bastani_generative_2025, kosmyna_your_2025}, and raises concerns regarding plagiarism and academic integrity \cite{westfall_educators_2023, kasneci_chatgpt_2023, bouteraa_understanding_2024, cotton_chatting_2024, playfoot_hey_2024}.

Therefore, instructors and researchers have raised urgent questions regarding how these AI tools should be integrated into higher education to mitigate dependency, 
with some argue that LLMs should be restricted from directly generating answers in educational contexts \cite{liffiton_codehelp_2024, bastani_generative_2025}. Approaches that introduce guardrails and embed richer pedagogical strategies, such as Socratic questioning \cite{paul_thinkers_2019}, have been shown to stimulate student thinking and reduce reliance on ready-made solutions \cite{chowdhury_autotutor_2024, sheese_patterns_2024, petrovska_incorporating_2024}. Similar strategies were recently implemented by OpenAI in ChatGPT's "study mode" \footnote{\url{https://openai.com/index/chatgpt-study-mode/}}. %\cite{openai_introducing_2025}. 
Furthermore, researchers have begun exploring how to reframe LLMs from "cheating tools" into metacognitive scaffolds \cite{gregory_assessing_1994, dunlosky_metacognition_2008} that support reflective learning \cite{boud_reflection_2013}, thereby prompting students to actively reflect on what they have learned rather than simply copying answers \cite{borge_using_2024, tankelevitch_metacognitive_2024, prasad_self-regulated_2024}. Subsequent research demonstrates that LLMs can improve learning outcomes by generating metacognitive prompts \cite{neshaei_metacognition_2025, neshaei_mindmate_2025} - guidance that encourages student reflection - which reduce the likelihood of answer-copying and improve exam performance \cite{kumar_guiding_2024, kumar_supporting_2024}.

%随着学生逐渐掌握LLMs的使用方式，LLMs已经被他们逐渐应用于高等教育学习中的多个部分\cite{baig_chatgpt_2024, budhiraja_its_2024}。由于当前关于使用LLMs的学生行为和教师规范的文献要么依赖于广泛的调查\cite{baig_chatgpt_2024, wu_systematic_2025, prather_robots_2023, ghimire_generative_2024, simkute_new_2025}，要么针对单一场景如写作 \cite{cotton_chatting_2024, budhiraja_its_2024, zheng_charting_2024} 或编程 \cite{amoozadeh_student-ai_2024, liffiton_codehelp_2024, sheese_patterns_2024, becker_programming_2023}的学生实践，以及这两个场景下的教师规范\cite{lau_ban_2023, barrett_not_2023, cotton_chatting_2024, prather_robots_2023, sheard_instructor_2024, park_promise_2024}。因此，以往的工作未能捕捉到当今学生使用LLMs的方式发生的微妙且依赖于具体情境的协商过程,即如何从简单的复制粘贴发展到更复杂的工作流程（例如，如何使用 LLM 完成基于项目的学习）以及教师规范如何变化以适应（或未能适应）这些全新行为的理解.The critical question is no longer a binary judgment of whether LLM usage is beneficial or detrimental, but rather how the boundaries of acceptable use are being redrawn across different pedagogical contexts. Furthermore, while implementing guardrails has proven effective in mathematics and programming \cite{bastani_generative_2025, liffiton_codehelp_2024, sheese_patterns_2024}, it remains unclear whether such designs can accommodate the more complex behaviors students now exhibit. These gaps collectively motivate our empirical study, which investigates the evolution of student behaviors and instructor norms to inform the future design of educational LLMs.

As students gain proficiency with LLMs, they have increasingly integrated these tools into diverse aspects of higher education learning \cite{baig_chatgpt_2024, budhiraja_its_2024}. Current literature on student practices and instructor norms, however, relies primarily on broad surveys \cite{baig_chatgpt_2024, wu_systematic_2025, prather_robots_2023, ghimire_generative_2024, simkute_new_2025} or studies confined to single domains, such as writing \cite{cotton_chatting_2024, budhiraja_its_2024, barrett_not_2023} or programming \cite{amoozadeh_student-ai_2024, liffiton_codehelp_2024, sheese_patterns_2024, becker_programming_2023}, along with the corresponding instructor perspectives \cite{lau_ban_2023, barrett_not_2023, cotton_chatting_2024, prather_robots_2023, sheard_instructor_2024, park_promise_2024}. Consequently, prior work fails to capture the nuanced, context-dependent negotiations that characterize how student practices evolve from simple copy-pasting to complex workflows and how instructor norms adapt (or fail to adapt) to these emerging behaviors. The critical question is no longer a binary judgment of whether LLM usage is beneficial or detrimental, but rather how the boundaries of acceptable use are being redrawn across different pedagogical contexts. Furthermore, while implementing guardrails has proven effective in mathematics and programming \cite{chowdhury_autotutor_2024, bastani_generative_2025, liffiton_codehelp_2024, sheese_patterns_2024}, it remains unclear whether such designs can accommodate the more complex behaviors students now exhibit in different scenarios. These gaps collectively motivate our empirical study, which investigates the evolution of student practices and instructor norms to inform the future design of educational LLMs.

%我们先后与来自计算机科学的16名学生和6名老师进行研讨会。我们总结了计算机科学教育中不同场景下的LLMs用例，归纳了不同用例下的学生行为和老师规范，讨论了不同用例下师生的冲突，据此探索针对不同用例的教学策略创新和LLMs设计建议。具体来说，我们的贡献如下：1）我们收集了学生在五大场景"Writing", "Quiz", "Information retrieval", "Project-based learning", and "Programming"下真实的LLMs用例，为每个用例归纳了学生处理LLMs输出的行为和观点。2）老师们对于这些用例的规范，并根据学生行为和老师规范将这些用例整合为7大用例，按其中学生行为和老师规范的冲突等级分类为Clear conflict: "Writing-generation" and "(Programming) quiz-solving"；Partial conflict: "Project-based learning" and "Programming project-implementation"；Basic alignment: "Writing-revision \& ideation", "(Programming) quiz-correction" and "Information retrieval"。3）这些存在冲突的用例中老师们为实现他们的规范进行的教育策略创新。4）一系列针对这些用例中LLMs如何助学的设计建议。5）师生视角下LLMs对师生关系的变化。
To address this gap, we conducted user studies with 16 students and 6 instructors from Computer Science (CS). We mapped student LLM intents across scenarios in CS education, gathered student practices and instructor norms for each intent, examined student-instructor conflicts, and explored pedagogical innovations and LLM design recommendations tailored to specific intents. Specifically, our contributions are: 
1) We identified 5 scenarios for using LLMs: "Writing", "Quiz", "Programming", "Project-based learning", and "Information retrieval". We then collected 7 intents, in which complex student workflows were exhibited in "writing-generation", "programming project-implementation", and "PBL-ideation". 
2) The 7 intents are grouped by conflict level between student practices and instructor norms: Clear conflict - "Writing-generation" and "(Programming) quiz-solving"; Partial conflict - "Programming project-implementation" and "PBL-ideation"; Broad agreement - "Writing-revision \& ideation", "(Programming) quiz-correction" and "Info-query \& summary". 
3) We describe pedagogical adaptations where instructor norms shift from banning LLMs to accepting and assessing the usage process (e.g., grading prompts and logs). 
4) We propose design recommendations, advocating for default guardrails with game-like and empathetic strategies to prevent students from "deserting" models in clear conflict intents, particularly in "Writing-generation", and comprehension checks for agreed-upon intents.

\section{Related work}
We focus on higher education in computer science because it offers numerous LLM application scenarios, instructors and students are more familiar with LLMs, and there is already a solid research foundation \cite{singh_exploring_2023, shoufan_exploring_2023, prather_robots_2023, park_promise_2024, amoozadeh_trust_2024, stone_exploring_2024, hou_effects_2024, margulieux_self-regulation_2024, sheard_instructor_2024, mahon_guidelines_2024, petrovska_incorporating_2024}. Related work are discussed in detail below.

%暂时不用
%LLM对计算机教育综述\cite{prather_robots_2023}
%Becker等人强调LLM使入门编程的教育更加简单，并认为以后编程老师的职责将转为代码阅读与评估\cite{becker_programming_2023}
%stone等人对LLM如何改变入门编程教育进行综述\cite{stone_exploring_2024}
%Yan et al.提出编程教育中可以引入LLM以提高协作编程的效率和结果,结果表明将LLM纳入协作编程可以显著减轻学生的认知负荷，提高他们的计算思维能力。\cite{yan_llm-based_2025}
%Bilstrup等人构建机器学习模型，旨在在中等教育中更广泛地教授 AI 素养\cite{bilstrup_ml-machineorg_2024}
%Tan等人发现教师面临着巨大的信息差距，在探索 ChatGPT 的定制学习任务能力并确保其适合不同学习者的需求方面缺乏明确性，中学老师\cite{tan_more_2024}
%Parker等人研究了Performance of ChatGPT in assessment Generation，发现ChatGPT的成绩普遍优于普通学生，强调了LLM在本科评估中的有效性和适应性\cite{parker_graduate_2024}
%Santana-Mancilla等人提出学生通过计算机游戏的设计和评估学习HCI\cite{santana-mancilla_teaching_2019}
\subsection{Student practices with LLMs}

Since ChatGPT's debut, student learning practices in higher education have become intertwined with LLMs. From students perspective, LLMs are valued for providing hyper-personalized instruction and bolstering learning confidence and efficiency \cite{kasneci_chatgpt_2023, chan_students_2023, shoufan_exploring_2023, adel_chatgpt_2024, mogavi_chatgpt_2024}. However, their trust in and adoption of these tools are contingent upon specific scenarios, disciplines, and individual factors \cite{kubullek_understanding_2024, amoozadeh_trust_2024, johnston_student_2024}, leading to calls for universities to provide formal guidance on effective usage \cite{singh_exploring_2023, shoufan_exploring_2023}. Objectively, meta-analyses confirm that LLMs can enhance academic performance and higher-order thinking, provided the pedagogical design is sound \cite{youssef_examining_2024, wang_effect_2025, kestin_ai_2025, deng_does_2025, baig_chatgpt_2024, wu_systematic_2025, park_promise_2024, nathaniel_investigating_2025}. Conversely, because AI-generated text and code are often sufficient to pass standard assessments \cite{malinka_educational_2023, playfoot_hey_2024, cotton_chatting_2024}, a reliance has emerged: students prioritizing course completion over deep learning increasingly depend on these tools. This dependency results in diminished critical thinking and poorer academic outcomes, while rising an academic integrity crisis \cite{kasneci_chatgpt_2023, cotton_chatting_2024, jost_impact_2024, bastani_generative_2025, rogers_chatgpts_2025, mogavi_chatgpt_2024, stadler_cognitive_2024, youssef_examining_2024, lee_impact_2025, kosmyna_your_2025}. Various indications suggest that how students use LLMs is related to their self-efficacy \cite{margulieux_self-regulation_2024}, exacerbating the polarization of student performance \cite{prather_widening_2024}. 

Existing literature has extensively mapped the general landscape of these practices through broad surveys, revealing a spectrum of usage ranging from minimal interaction to "all-purpose" heavy reliance \cite{stojanov_university_2024}, that is, students use LLMs as on-demand partners for diverse tasks, including ideation, information retrieval, writing, programming, and even research \cite{stojanov_university_2024, budhiraja_its_2024, rajabi_unleashing_2024, hou_effects_2024, amoozadeh_trust_2024, simkute_new_2025}. 
In terms of scenario-specific practices, the literature is saturated with studies on writing and programming. In writing, research has detailed how students use LLMs for brainstorming, outlining, drafting, revising, feedback, and evaluation \cite{barrett_not_2023}. However, students prefer utilizing LLMs for direct drafting and revision to improve efficiency \cite{barrett_not_2023, kong_pedagogical_2024}. This practice - often accompanied by manual obfuscation of AI-generated text to evade detection  \cite{adnin_examining_2025} - has been shown to degrade writing quality \cite{stadler_cognitive_2024, cotton_chatting_2024, wang_effect_2025}.
Similarly, in programming, concerns center on students' tendency to blindly copy code and debugging solutions \cite{becker_programming_2023, liffiton_codehelp_2024, amoozadeh_student-ai_2024}, whereas leveraging LLMs for code explanation, cross-verification, or task decomposition has a positive impact on programming learning \cite{becker_programming_2023, kazemitabaar_improving_2024, budhiraja_its_2024, nathaniel_investigating_2025}. Beyond these scenarios, while some students envision LLMs supporting project-based learning \cite{zheng_charting_2024}, the actual workflows remain opaque.

As LLMs becomes ubiquitous, students may be developing more complex strategies to integrate LLMs into their learning. While the mentioned studies indicate that students' LLM workflows are becoming increasingly complex, current research has only deeply explored writing and programming, leaving a critical gap in understanding student practices in other scenarios, such as project-based learning and information retrieval. Specifically, it remains unclear whether students across different scenarios are blindly copying generated content, partially relying on it, or using it for divergent thinking and transform LLM outputs into final submissions \cite{simkute_new_2025, baig_chatgpt_2024, wu_systematic_2025, stone_exploring_2024, prather_robots_2023}. This motivates us to investigate how students translate LLM responses into learning outcomes across different pedagogical scenarios.
%随着LLM的使用变得无处不在，学生们可能正在发展更复杂的策略将LLMs融入他们的学习中.以上研究说明学生的LLMs工作流程正在日益复杂化，但目前研究仅对写作和编程作了深入探索，而对当前学生在不同场景下，如PBL和信息检索，的实践的理解存在一个关键缺口。具体而言，不同场景下学生们究竟是盲目复制生成的内容，还是部分依赖它，或是利用它来激发发散性思维，将 LLM 的输出转化为最终的提交，这些都尚不明确\cite{simkute_new_2025, baig_chatgpt_2024, wu_systematic_2025, stone_exploring_2024, prather_robots_2023}。这激励我们深入探究学生如何在不同的教学场景中将LLM回答转化为学习成果。

\subsection{Instructor views on LLMs for studying}
%教师们认为LLMs有改善教学的潜力，但对学生使用LLMs的方式存在分歧，这些研究局限在写作和编程场景。教师认为LLMs不应充当"答案生成器"，因为LLMs代学生写作和编程会破坏测评的公平性和效度 \cite{cotton_chatting_2024, prather_robots_2023, sheard_instructor_2024}。相较之下，教师认可在写作中将 LLMs 用于头脑风暴和修改润色 \cite{barrett_not_2023}，在编程中将其用于代码构思和错误纠正 \cite{sheard_instructor_2024}。由于LLMs的普及和双刃剑性质，教师们放弃了单纯的"禁止"策略，转向寻求通过调整教学法将LLMs作为教学脚手架\cite{sheard_instructor_2024, lau_ban_2023, mahon_guidelines_2024, ghimire_generative_2024}，并已经形成初步共识：1）在作业中披露与标注 LLM 使用 \cite{adnin_examining_2025, cotton_chatting_2024}; 2）课程评估侧重线下考试，作业评估纳入过程日志和相关解释 \cite{prather_robots_2023, sheard_instructor_2024, lau_ban_2023, cotton_chatting_2024, zheng_charting_2024, denny_explaining_2024, denny_prompt_2024}；3）设立课程来提升学生 AI 素养与批判思维 \cite{kasneci_chatgpt_2023, cao_ai_2025, prabhudesai_here_2025, wu_systematic_2025}。但研究者和教师们也指出，不同场景下教师与学生对于 LLMs 使用的可接受边界仍非常模糊且十分必要\cite{barrett_not_2023, rajabi_unleashing_2024, park_promise_2024, mahon_guidelines_2024}。这也意味着我们仍不清楚当下教师的规范如何适应对学生更复杂的使用LLMs的方式，以及这些规范如何影响教师进一步调整教学法，和对LLMs设计的思考。了解这些空白有助于高校和教育机构制定清晰的指南、政策和课程，以确保LLMs负责任的用于教学目的\cite{harvey_dont_2025, hasanein_drivers_2023, shoufan_exploring_2023}，帮助教育社区通过教学法，工具设计和政策将LLMs从一把"双刃剑"转变为"学习伙伴"\cite{wu_systematic_2025, park_promise_2024}。

Instructors recognize the potential of LLMs to enhance instruction, yet remain divided regarding student usage patterns - a discourse confined to writing and programming scenarios. Instructors contend that LLMs should not serve as "answer generators", arguing that having LLMs write or code on behalf of students undermines the fairness and validity of assessments \cite{cotton_chatting_2024, prather_robots_2023, sheard_instructor_2024}. In contrast, instructors approve of utilizing LLMs for brainstorming and polishing in writing \cite{barrett_not_2023}, as well as for conceptualizing code and debugging in programming \cite{sheard_instructor_2024}. Given the ubiquity and double-edged nature of LLMs, instructors have moved away from strict "prohibition", instead adapting pedagogics to integrate LLMs as educational scaffolds \cite{sheard_instructor_2024, lau_ban_2023, mahon_guidelines_2024, ghimire_generative_2024}. A preliminary consensus has emerged focusing on: 1) requiring the disclosure and citation of LLM usage in assignments \cite{adnin_examining_2025, cotton_chatting_2024}; 2) prioritizing offline examinations for course assessment while incorporating process logs and explanation into assignment evaluation \cite{prather_robots_2023, sheard_instructor_2024, lau_ban_2023, cotton_chatting_2024, zheng_charting_2024, denny_explaining_2024, denny_prompt_2024}; and 3) establishing curricula to enhance students' AI literacy and critical thinking \cite{kasneci_chatgpt_2023, cao_ai_2025, prabhudesai_here_2025, wu_systematic_2025}.

However, researchers and instructors note that the boundaries of acceptable LLM usage across different scenarios remain ambiguous, yet defining them is critical \cite{barrett_not_2023, rajabi_unleashing_2024, park_promise_2024, mahon_guidelines_2024}. This implies that it remains unclear how current instructor norms adapt to increasingly complex student usage patterns, or how these norms influence further pedagogical adjustments and reflections on LLM design. Addressing these gaps will assist universities and educational institutions in formulating clear guidelines, policies, and curricula to ensure the responsible use of LLMs for educational purposes \cite{harvey_dont_2025, hasanein_drivers_2023, shoufan_exploring_2023}, ultimately helping the educational community transform LLMs from a "double-edged sword" into a "learning partner" through pedagogical innovation, tool design, and policy adaptation \cite{wu_systematic_2025, park_promise_2024}.

\subsection{Interaction designs for learning with LLMs}

%研究者们试图通过人机交互技术创新来避免学生对LLMs的依赖，已经取得初步进展。Bastani et al.\cite{bastani_generative_2025}通过大规模调查提出如果LLMs不添加护栏以禁止生成可用答案会损害学习能力。但添加护栏后，LLMs应该生成什么内容来帮助学生学习？元认知指导作为教师引导学生反思的主要方式受到关注。起初，研究者表明人机协作交互有望支持学生的元认知训练\cite{borge_using_2024}，且提供个性化的反思性学习\cite{yuan_generative_2024}，随后提出面向生成式AI的元认知设计要点\cite{tankelevitch_metacognitive_2024}. 接着研究者开始实证研究LLMs作为元认知脚手架的影响，Kumar et al.\cite{kumar_guiding_2024, kumar_supporting_2024}发现将LLMs用来教学反思策略显著抑制了学生抄答案的行为并提高了课程成绩。Neshaei et al. \cite{neshaei_metacognition_2025,neshaei_mindmate_2025}表明LLMs可以支持反思性写作。此外，LLMs还可以通过不同设计来提高学生的元认知和批判性思维，如允许学生比较不同回答\cite{prabhudesai_here_2025}，引入认知冲突\cite{akmam_integration_2024}，或支持学生自我监控学习过程\cite{prasad_self-regulated_2024, chaudhury_milestones_2025}. 因此，为LLMs添加护栏转而生成元认知指导，如苏格拉底反问和提示，是避免学生依赖LLMs并提高学习能力的主要方法\cite{chowdhury_autotutor_2024, bastani_generative_2025}。该策略也被Liffiton 和Sheese et al.应用于编程教学，证明其显著降低了学生使用LLMs编码的依赖，表明该工具的使用频次与成绩呈正相关。但这一策略在被OpenAI实现为ChatGPT的"study mode"后在实际使用中被质疑，即学生可以且通常直接使用普通模式，导致护栏无用\cite{rogers_chatgpts_2025}。而且该策略仅在数学和编程场景被证实有效，但我们对其是否在其他场景，特别是写作这种学生会直接复制生成文章的场景起作用仍一无所知。为此，我们将在用户研究中探索学生和教师对这一策略的看法。
Researchers have begun mitigating student over-reliance on LLMs through HCI innovations. Bastani et al. \cite{bastani_generative_2025}, through a large-scale survey, argued that LLMs can harm learning if they generate directly usable answers without guardrails. However, with guardrails in place, the key question becomes: kinds of outputs best support learning? Metacognitive guidance \cite{gregory_assessing_1994, dunlosky_metacognition_2008} - widely used by instructors to induce reflection - has emerged as a key solution. Early work demonstrated the potential of human-computer collaboration to support metacognitive training \cite{borge_using_2024} and personalized reflective learning \cite{yuan_generative_2024}, leading to design principles for metacognition in generative AI \cite{tankelevitch_metacognitive_2024}. Subsequent studies examined LLMs as metacognitive scaffolds: Kumar et al. \cite{kumar_guiding_2024, kumar_supporting_2024} found that teaching reflection strategies with LLMs reduced answer-copying and improved performance, while Neshaei et al. \cite{neshaei_metacognition_2025, neshaei_mindmate_2025} showed that LLMs can effectively support reflective writing. Additional design interventions include enabling answer comparison \cite{prabhudesai_here_2025}, introducing cognitive conflict \cite{akmam_integration_2024}, and supporting self-regulated learning \cite{prasad_self-regulated_2024, chaudhury_milestones_2025}.

Consequently, integrating guardrails that redirect LLMs to generate metacognitive guidance, such as Socratic questioning \cite{paul_thinkers_2019} and reflective prompts, has become a primary strategy to prevent over-reliance and boost learning outcomes \cite{chowdhury_autotutor_2024, petrovska_incorporating_2024, bastani_generative_2025}. In programming education, Liffiton \cite{liffiton_codehelp_2024} and Sheese et al. \cite{sheese_patterns_2024} showed that such designs reduce dependence on code generation and correlate tool use with better performance. Yet the real-world effectiveness of this strategy - recently instantiated as OpenAI’s "study mode" - is uncertain, as students can and often bypass it by switching back to standard mode \cite{rogers_chatgpts_2025}. Moreover, although effective in mathematics \cite{chowdhury_autotutor_2024, bastani_generative_2025} and programming \cite{liffiton_codehelp_2024, sheese_patterns_2024}, its applicability to other scenarios remains unclear, especially in writing, where students often copy generated text verbatim. To address this gap, we explore student and instructor perceptions of this strategy in our user study.

\section{Research approach}

We conducted two linked user studies: one with students and one with instructors. We first ran one-on-one sessions with CS students to collect and synthesize their practices and perspectives on handling LLM outputs across different intents situated in varied scenarios (a scenario may have several intents). We then invited CS instructors with teaching experience to review student practices, eliciting instructor norms, related pedagogical strategies, and LLM design insights. The study received ethical approval from [anonymized]. Below, we detail the student user study, and then describe the instructor user study (Section \ref{S:M-Instructors}) after reporting findings from the student user study. Through user studies with students and instructors, we address the following research questions: RQ1) What are the scenarios and intents for CS students using LLMs in education? For each intent, how do they handle LLM outputs (copy results or support learning)? RQ2) What norms do CS instructors set for LLMs, and do these conflict with student practices in RQ1? RQ3) What changes are taking place in student practices and instructor norms since LLMs emerged? RQ4) How are these changes transforming pedagogical strategies, LLM design?
%, and instructor–student relationships?

\section{Method - Students}

\subsection{Participants}
\begin{table*}[htbp]
\centering
\caption{Participants' IDs, Age, Gender (M = Male, F = Female), Education [UG = undergraduate (UG 1--2 = years 1--2; UG 3--4 = years 3--4)], LLM exp = experience using LLMs (1 = Unfamiliar to 5 = Expert), Number of chat logs, Scenarios = scenarios of chat logs (Info retrieval = Information retrieval, PBL = Project-based learning).}
\label{tab:students}
\begin{tabular}{lcccccl}
\toprule
ID & Age & Gender & Edu & LLM exp & Number of intents  & Scenarios \\
\midrule
P1 & 24 & F & Master & 3 & 2 & Writing, Info retrieval \\
P2 & 25 & M & PhD & 3 & 2 & Writing, Programming \\
P3 & 19 & M & UG 1-2 & 3 & 4 & Quiz, Info retrieval, PBL, Programming \\
P4 & 23 & M & Master & 2 & 3 & Writing, Quiz, Info retrieval \\
P5 & 22 & M & UG 3-4 & 3 & 3 & Writing, Info retrieval, Programming \\
P6 & 20 & M & UG 1-2 & 4 & 5 & Writing*2, Quiz, Info retrieval, Programming \\
P7 & 22 & F & UG 3-4 & 3 & 3 & Writing, PBL, Programming \\
P8 & 23 & F & Master & 3 & 3 & Writing, PBL, Programming \\
P9 & 24 & M & Master & 2 & 3 & Writing, Info retrieval, Programming \\
P10 & 19 & M & UG 1-2 & 4 & 3 & Writing, Quiz, Programming \\
P11 & 19 & F & UG 1-2 & 3 & 3 & Quiz, Info retrieval, Programming \\
P12 & 21 & M & UG 3-4 & 4 & 3 & Quiz, Info retrieval, Programming \\
P13 & 25 & F & PhD & 5 & 3 & Writing, Info retrieval*2 \\
P14 & 25 & M & PhD & 5 & 3 & Writing, Info retrieval, Programming \\
P15 & 25 & M & Master & 4 & 4 & Info retrieval, PBL*2, Programming \\
P16 & 25 & M & PhD & 5 & 3 & Info retrieval, PBL, Programming \\
\bottomrule
\end{tabular}
\end{table*}
%我们通过大学邮件系统和社交媒体邀请最终确定招募了总共16名计算机科学的学生作为参与者。参与者被要求完成一份简单的人口统计问卷，涵盖了年龄，性别，受教育程度（在读年级），LLMs使用经验（使用五级李克特量表），展示在Table \ref{}。This demographic information helped target inquiries within our study. The participants' ages ranged from 19 to 25 (mean=22.5, SD=2.3).其中11名自称为男性，另外5名为女性。我们确保了参与者来自高等教育的不同受教育水平，包含4个1-2年级本科生，3个3-4年级本科生，5个研究生与4个博士生。这使我们能涵盖计算机科学中不同教育程度下学生的实践和观点。此外，参与者们都确定自己有过将LLMs用于学习的经历，且普遍拥有较高的LLM使用经验（mean=3.5）。每个学生参与者在完成session后都将获得50元人民币作为补偿。
We recruited a total of 16 CS students via university email lists and social media. Participants were asked to first complete a brief demographic survey covering age, gender, education level (current year), and LLM usage experience (five-point Likert scale \cite{joshi_likert_2015}), as shown in Table \ref{tab:students}. This demographic information helped target inquiries within our study. The participants' ages ranged from 19 to 25 (mean = 22.5, SD = 2.3). Of these, 11 identified as male and 5 as female. We ensured that participants represented multiple levels of higher education, including 4 undergraduates in years 1--2, 3 undergraduates in years 3--4, 5 master's students, and 4 PhD students. This allowed us to cover practices and perspectives across different educational stages in CS. In addition, all participants confirmed that they had previously used LLMs for learning and generally reported high LLM experience (mean = 3.5). Each student participant received CNY 50 as compensation upon completing the session.

\subsection{User study procedure}

The first author conducted one-on-one sessions with each student participant, totaling 16 sessions by the time we completed the study, with each session lasting approximately 1.5 hours. Participants could choose to attend the sessions either in person or remotely. The detailed procedure of the session and the interview questions are as follows.

We began by asking participants to list the \textbf{scenarios (e.g., writing, programming)} in which they use LLMs and to describe their \textbf{intents (e.g., generating writing, solving programming quizzes)} (usually 1–-3) for each scenario. We then moved to the subsequent phase - chat logs analysis. To ensure authenticity, we required participants to access their recent LLM chat logs covering the intents they described. They then copied the prompts and LLM outputs (potentially multi-turn) into a Word document and shared it with us. \textbf{Each chat log corresponded to one intent and to a Word document.} Since participants used different LLMs, the chat log sources included ChatGPT (GPT-5), Gemini-2.5 \& 3, DeepSeek-R1, Kimi-K2, Qwen-3, and Doubao.

\subsubsection{Chat logs analysis}
In this phase, for every chat log:
\begin{enumerate}
  \item Students used Word's commenting function to highlight and annotate how they acted on parts of the LLM outputs they regarded as directly usable results or answers (e.g., verbatim or selective copying with minor edits) or treated as learning supports (e.g., seeking inspiration and stimulating their own thinking).
  \item For each annotation, students were also asked to explain the rationale for their behavior and whether there were any follow-up actions. For example, if annotated as copying: why copy? Did they submit the assignment after partial edits? If annotated as learning support: why use it for support? Did they cross-validate by other means? Did new questions arise that led them to continue querying the LLM?
\end{enumerate}

After annotating a chat log, participants were interviewed with the questions below:
\begin{enumerate}
  \item In this intent, do you generally prefer to copy LLM outputs or to use them only as learning support? When would you switch modes, and why?
  \item What role should the LLM primarily play here (e.g., answer provider, hinting tutor, brainstorming partner, code explainer, error spotter)?
  \item Would you seek help from an instructor in this intent? If yes/no, why? If yes, what do instructors typically do that helps?
  \item If you prefer copying answers: Would you accept an LLM that withholds answers and instead scaffolds learning? Why or why not?
  \begin{enumerate}
      \item Design suggestion: What specific design changes would make that acceptable?
  \end{enumerate}
  \item If you prefer learning support: Does receiving answers from LLMs benefit your learning?
  \begin{enumerate}
      \item Design suggestion: What additional scaffolds or interaction features would make the LLM's assistance more effective in this intent?
  \end{enumerate}
\end{enumerate}

This structured annotation, plus intent-based reflection, provided both fine-grained behavioral evidence (how LLM outputs were actually used) and corresponding rationales (why that practice made sense in the intent), enabling us to summarize changes in student practices across intents in different scenarios.

\subsubsection{Ethical discussion}
%由于LLMs被认为会生成幻觉等不负责任的内容，这些不负责任的内容可能对学生学习造成负面影响，因此我们需要了解学生在面对生成的结果时是否受到这些内容的干扰。此外，有老师担忧当学生逐渐倾向于使用LLMs学习后，老师的职能可能受到威胁，我们已经在chat logs analysis phase中让学生表达了每个场景下他们是否会向老师求助，我们还希望据此看看学生按总体来说对此有什么见解。询问这些问题helping us connect scenario-specific practices to broader ethical and pedagogical implications for LLM-supported learning. 因此，After completing all intents, participants engaged in a focused discussion on 负责任的LLM and role boundaries
Since LLMs may generate irresponsible contents like hallucinations that may negatively affect learning \cite{kasneci_chatgpt_2023, mogavi_chatgpt_2024, park_promise_2024, wu_systematic_2025}, we asked whether students are harmed by such outputs. Moreover, given concerns that growing LLM use may threaten instructor roles \cite{harvey_dont_2025, hasanein_drivers_2023, barrett_not_2023, prather_robots_2023}, we not only had students indicate whether they seek instructor help in each intent but also examined their overall views on this issue. Asking these questions helps us link intent-specific practices to broader ethical and pedagogical implications for LLMs. Therefore, student participants finally engaged in a discussion on responsible LLM use:
\begin{enumerate}
  \item Did you encounter irresponsible responses (e.g., hallucinations or biased advice) in these intents?
  \item If LLMs cannot provide answers and instead guide thinking or offer strategies that help reflect, would this reduce such irresponsible outputs, or might it generate new hallucinations?
  \item In which intents do you feel the LLM replaces aspects of the instructor's role? Conversely, which instructor functions are not replaceable by LLMs?
\end{enumerate}

\subsection{Data analysis}\label{S:DataA-Students}
%All sessions were recorded and transcribed%，所有用例对应的chat logs between the participants and LLMs，包含学生的注释和summary，都被保存.我们最终收集了51份chat logs，以及16份transcripts containing each participant's think-aloud and interview data。For each session, we conducting thematic analysis \cite{braun_using_2006} of chat logs and the transcript. In our initial coding round, we coded 所有chat logs所属场景并categorized them into "Writing"，"Problem-solving"，"Information retrieval"，"Project-based learning (PBL)"，"Coding".在第二轮编码中，我们对每个场景下的chat logs所属用例进行编码，并分类为"Writing"： "生成", "修改"和"启发"；"Problem-solving"没用分类用例；"Information retrieval"："概念查询"和"信息摘要"；"Project-based learning (PBL)"："research", "experiments", and "engineering projects"；"Coding"："Coding problems"和"Coding projects"。最后，在第三轮编码，we coded statements related to our research questions并将他们分类为RQ1："行为"和"观点"；RQ4："LLM设计建议"。Table 展示了每个参与者贡献的chat logs （用例）数量以及所属的场景。我们将在接下来的section \ref{S:F-Students}按场景分类来详细介绍我们的发现。

All sessions were recorded and transcribed, all chat logs between the participants and LLMs, including student annotations, were saved. We ultimately collected 50 chat logs and 16 transcripts containing each participant's interview data. For each session, the first author independently conducted thematic analysis \cite{braun_using_2006} of chat logs and the transcript. In our initial coding round, we coded the scenario of each chat log and categorized them into "Writing", "Quiz", "Programming", "Project-based learning", and "Information retrieval". In the second coding round, we coded the intents within each scenario and organized them as follows. For "Writing": "Writing-generation", "Writing-revision", and "Writing-ideation"; for "Quiz": "Quiz-solving" and "Quiz-correction"; for "Programming": "Programming quiz-solving", "Programming quiz-correction", and "Programming projects-implementation"; for "Project-based learning (PBL)": "PBL-ideation"; and for "Information retrieval": "Info-query" and "Info-summary". Finally, in the third coding round, we coded statements related to our RQs within each intent and grouped them under RQ1: "Practices" and "Perspectives" and RQ4: "Design advice". Table \ref{tab:students} shows the number of chat logs contributed by each participant and their associated scenarios. We present findings in Section \ref{S:F-Students} based on intents.
%; RQ3: "Instructor or LLM";

\section{Findings: Student intents and practices}\label{S:F-Students}
We answer RQ1 in this section by presenting student practices and perspectives across different intents under five scenarios in handling LLM outputs. We place ethical discussion regarding irresponsible outputs in the appendix \ref{A:Ethical} as it is not related to paper's main contributions.
\subsection{Writing}

This scenario involves 13 chat logs from 12 students, covering three intents: Writing-generation, Writing-revision, and Writing-ideation (P6 provided chat logs for both "Writing-ideation" and "Writing-generation").

In CS courses, this scenario typically entails students writing literature reviews, lab reports, and papers. Eleven students stated that they primarily used LLMs for "Writing-generation" and provided corresponding chat logs. Only P10 indicated that they would not copy LLM-generated articles, but rather use them as a reference to conceptualize their own essays, noting that "LLM-generated writings contain a lot of fluff, so I don't dare use them directly for assignments" (P10). Additionally, two students demonstrated chat logs reflecting other intents. First, regarding "Writing-revision", P2 stated that they primarily used LLMs to modify (including translate and polish) their writing, comparing the versions before and after to decide whether to adopt the changes. This intent was also mentioned by several other students. P2 explained that this comparison is necessary because LLM modifications do not always align with their original ideas and may suffer from faithfulness hallucinations. P2 emphasized that obtaining a refined result is the primary goal. Second, regarding "Writing-ideation", P6 indicated using LLMs to ideate and expand writing directions. This is because assignment requirements are sometimes limited to a simple topic, such as "a review of the field of artificial intelligence", causing P6 (a first-year undergraduate) to struggle with topic selection. Under this intent, P6 only draws inspiration from LLM outputs. These findings echo the research of Barrett et al. \cite{barrett_not_2023}.

However, we observed more nuanced student practices under the "Writing-generation" intent. Generally, students tended to copy LLM-generated writing, though their specific methods varied. Six students (P1/4/5/9/13/14) reported inputting writing requirements, copying the generated writing, and making minor edits. Three other students (P6, 8, 12) emphasized that outlines enable LLMs to generate articles that better meet their needs, demonstrating an "outline $\rightarrow$ essay" workflow. P8 first drafted a sectioned outline with descriptions, then used the LLM to expand it to the required word count. P6 and P12 asked the LLM to first generate outlines, revised it themselves, and then fed the revised outline back for full-essay generation. All three typically copied the result with minor edits. They also explained why outline-first: P6 and P8 found that generating a full essay directly caused more issues and required more manual fixes, while P12 regarded this workflow as a comprehensive prompt engineering that yielded more satisfactory results. 

Two students shared alternative practices. To complete lab reports, P5 fed the experiment objectives, algorithm code, datasets, and results into the LLM to generate the report, finally copying the result with minor edits. Across these cases, "minor edits" included fixing misalignments with requirements, personal contexts, or experimental settings (P1/4/7/14); deleting overly long portions (P1) or asking LLMs for expansion when content was insufficient (P4/5/7/13); and correcting hallucinations such as incorrect data (P13) or erroneous citations (P6/13). Students refrained from blind copying to avoid low grades caused by non-compliance with requirements or high AI detection rates. P7 employed multiple LLMs with the same prompt to generate the essay, explaining: "One LLM might write a sentence more logically, while another writes a paragraph better, and then I synthesize the content from different LLMs into the final article. This method also slightly reduces the AI detection rate. We summarize these practices as follows: using three generation approaches - "Direct generation", "Outline $\rightarrow$ essay generation", and "Material-based generation" - followed by copying with minor edits, or employing "Cross-model synthesis".

Students explained their motivations of using LLMs for writing: ten cited efficiency; three (P7/8/14) felt LLM writing was superior to their own; and two (P8/14) wanted to minimize thinking. Within the "Writing-generation" intent, seven students did not want LLMs to withhold full-essay generation, as using LLMs to finish the writing was precisely their goal; three (P4/5/9) explicitly stated they would switch to another LLM if one refused to generate essays. Interestingly, the three outline-first students (P6/8/12) were open to guidance rather than direct essay generation, though P8 limited this preference to high-stakes cases (e.g., publications). Additionally, P10, due to personal preference, favored LLMs that aid in thinking over those that provide complete texts.

\subsection{Quiz}
%In scenario "Quiz", a total of 6 students provided 6 chat logs，涵盖两个意图. "Quiz-solving" refers to students using LLMs to complete course quizzes, including calculation questions (P4), analytical questions (P3/11), and proof tasks (P6/10), totally 5 students. "Quiz-correction" refers to students using LLMs to correct quiz的errors, such as correcting calculation questions (P12).

%对于"Quiz-solving"意图，学生的实践主要取决于deadline。在deadline临近时，四名学生（P6除外）明确表示会直接复制LLM生成的答案，且提交后不会重新独立完成。此外，P3还说"如果测验内容与考试挂钩，我会主动学习；否则我就直接抄答案"。这种实践与多个研究发现一致，被证明损害学习表现，应该为LLMs添加护栏以避免这类实践\cite{bastani_generative_2025, chowdhury_autotutor_2024}。而我们还发现，五名学生平时更倾向利用LLM生成答案以学习解题逻辑。这不仅是因为考试无法使用LLMs，还因为LLM的推理过程和答案并不总是正确的（P3/4/11）。当时间充裕时，他们会阅读并验证推理过程，或者询问LLM"为什么答案不是[...]"（P11）。另外，仅P12提供了"Quiz-correction"的chat log：该学生要求LLMs解释其计算中的错误并生成正确答案，以学习解题思路。因此我们进一步探讨了该场景下学生对护栏的看法。所有学生都强调这与自我能动性有关，他们并不希望在他们想要获取答案时LLMs存在护栏，特别在"Quiz-correction"意图下获取答案本身就是学习过程。其中两人（P4/10）表示，若在时间紧迫时遇到LLM拒绝提供答案，他们会转而使用其他LLM。关于护栏是否有助于学习，（P3/12）表示，若回复中同时包含推理和答案，会降低他们自主尝试的意愿，并削弱思考的深度。相反，P11认为回答中同时包含推理和答案有助于学习；P10则认为"先推理后给出答案"的方式会略微降低效率。

In scenario "Quiz", a total of 6 students provided 6 chat logs spanning two intents. "Quiz-solving" refers to students using LLMs to complete course quizzes, including calculation questions (P4), analytical questions (P3/11), and proof tasks (P6/10), totally 5 students. "Quiz-correction" refers to students using LLMs to correct errors in their quizzes, such as correcting calculation mistakes (P12).

For the "Quiz-solving" intent, student practices depends on deadlines. When approaching a deadline, four students (excluding P6) admitted to copying answers from LLMs without revisiting after submission. P3 further noted, "If the quiz is linked to exams, I will actively study it; otherwise, I just copy the answer". This reliance aligns with prior findings that such practice harms learning, supporting calls for guardrails \cite{bastani_generative_2025, chowdhury_autotutor_2024}. Conversely, with sufficient time, five students prefer understanding problem-solving logic from LLM-generated answer. Not only due to closed-book exams but also they should verify potential hallucinations in LLM reasoning and answers (P3/4/11) or ask "Why is the answer not [...]?" (P11). Additionally, only P12 provided a chat log for "Quiz-correction", where the student requested the LLM to explain calculation errors and generate the correct solution to understand the methodology.

Regarding guardrails, all students emphasized self-agency, opposing restrictions on answer generation - especially for "Quiz-correction" where the correct answer is integral to learning. Two students (P4/10) indicated they would switch models if refused. Opinions on whether guardrails facilitate learning varied, P3 and P12 noted that providing both reasoning and final answers reduces their willingness to attempt problems and discourages independent thinking. Whereas P11 argued including both aids learning, and P10 felt that a "reasoning-first, answer-second" format slightly reduced efficiency.

\subsection{Programming}
This scenario involves 13 chat logs from 13 students. Given that students distinguish the intents behind programming quizzes and projects, we categorize these into two sub-scenarios encompassing three specific intents: "Programming quiz-solving" (P9/10/16), "Programming quiz-correction" (P3), and "Programming project-implementation" (9 students).

\subsubsection{Programming quizzes}
Programming quizzes are programming languages (e.g., C, Java) or algorithmic interview questions requiring functionally correct code per requirements to implement specific logic. Three students contributed: algorithmic interview questions (P9/16) and course assignment quizzes (P10). Student practices and views in "Programming quiz-solving" mirrored "Quiz-solving": P10 would copy code to finish assignments before the deadline; all three preferred using LLMs to learn coding approaches since LLMs can't be used in exams or interviews. Under normal circumstances, all three focus on understanding the code logic, then attempt manual implementation. Three students differed in their attitudes toward receiving answers: P9 noted LLM-generated code with comments aids understanding more than explanations of logic without answer; P10 preferred outputting both code and logic to compare with manual work and probe mistakes; by contrast, P16 valued logic explanations over code. "Programming quiz-correction" covers correcting errors in programming quizzes. P3 first implemented code, then used LLMs to analyze errors after a failed run. This aligns with "Quiz-correction", again because LLMs cannot be used in programming exams. Since they use LLMs for debugging, P3 believes the LLM must provide correct code. In this sub-scenario, students’ use of LLMs to copy, explain, and debug code is consistent with the findings of most programming education research \cite{becker_programming_2023, liffiton_codehelp_2024, amoozadeh_student-ai_2024, kazemitabaar_improving_2024, budhiraja_its_2024, nathaniel_investigating_2025, shoufan_exploring_2023}.

\subsubsection{Programming projects}

Programming projects involve complete, functionally complex implementations (e.g., algorithm experiments, web apps). However, prior research focuses predominantly on programming quizzes, often overlooking this sub-scenario. Here, the primary intent is "Programming project-implementation", encompassing diverse practices. Students used LLMs for simple tasks like plotting (P2) and file edits (P5); and complex ones like modularizing files (P8), compiler configuration (P11), data analysis (P14), debugging (P7/15), and lab code implementation (P6/12). Six students copied generated code directly, adjusting only parameters like file paths. If execution failed, they iterated with the LLM using error messages until successful - a pattern observed in both simple (P2/P5) and complex tasks (P8/11/14/15). Their goal was complete implementation rather than understanding code mechanics. However, since LLMs rarely generate compliant, runnable code in a single turn, most students followed a "Copy $\rightarrow$ Run $\rightarrow$ Error $\rightarrow$ Fix" iterative workflow.

In contrast, three students (P6/7/12) engaged in selective copying followed by independent implementation to finish coding. For instance, P6 copied simple components (e.g., data loading) for a machine learning framework but implemented complex sections (e.g., pre-processing, training) himself. P12 debugged copied code to understand parameter functions, while P7 learned fixes through LLM error localization. Overall, six students (P2/5/8/11/14/15) relied on the "Copy $\rightarrow$ Run $\rightarrow$ Error $\rightarrow$ Fix" iterative workflow to complete programming projects, whereas three (P6/7/12) explicitly used LLMs for learning. Notably, three (P2/5/6) copied simple functional code without studying. Several students prioritizing completion (P2/5/8/11/14) opposed guardrails, with three (P2/5/14) stating they would abandon LLMs if code generation are restricted.

\subsection{Project-based learning}
%学生将LLMs用于工程项目，研究，实验的规划设计被归于基于项目的学习（PBL）场景，而该场景下通常没有固定的答案。该场景不分用例，总计5个学生提供6个聊天记录包括研究方案设计（P16），项目规划（P7，8，15），实验优化（P3，15）。
%大部分学生（P3，P8，P15，P16）不认为LLMs在该场景下的回答是完全可用的，他们主要把 LLMs 用作辅助思考的工具，在多轮交互中借助 LLM 生成和改进方案。在优化实验种，学生将LLM输出的改进电路方案（P3）和硬件加工方法（P15）作为参考，再结合实际需求进行修改。学生们还在项目规划与设计中将LLM用于发散思维，比如探索线上平台的功能改进方向再去寻找开源项目（P8），或是头脑风暴数据的可视化方案（P15），根据研究方向和个人偏好设计研究方法（P16）。学生中仅有P7在用LLM生成公司数据分析方案的时候直接复制了LLM建议的数据结构，这是因为她说当时并没有思路，但她也表示之后她在和老师与公司的交流以及完成实际分析方案的过程中不断修正了数据结构。学生们普遍认为LLMs可以作为具有丰富经验的研究者，实验专家和工程师等角色，基于其丰富的数据提出潜在可行的建议，为他们完成PBL提供丰富思路。
In this scenario, 5 students provided 6 chat logs, including research methods design (P16), engineering project planning (P7/8/15), and experiment optimization (P3/15). Since student practices and views were similar across these tasks, we name their intent under this scenario as "PBL-ideation".

Most students (P3/8/15/16) did not see LLM responses as fully usable in this intent; instead, they mainly treated LLMs as tools for supporting thinking, iteratively inspiring and refining plans through multi-turn interactions. 
In experiment, students used LLM-suggested improvements to circuit designs (P3) and hardware fabrication methods (P15) as references, then adapted them to practical needs. They also used LLMs for divergent ideation in research and engineering projects. For example, designing research methods based on research directions and personal preferences (P16); exploring ways to improve an online platform and then searching open-source projects (P8); brainstorming data visualization approaches (P15). We summarize this working process as an "Inspire $\rightarrow$ Contextually refine $\rightarrow$ Attempt" iterative workflow. Only P7 copied the LLM-generated data structure for a company data-analysis plan because they initially had no idea. But they subsequently revised it after discussions with the instructor and the company. Overall, students commonly viewed LLMs as experienced researchers, engineers, or lab experts that offer potentially feasible suggestions and provide rich ideas for project-based learning.

\subsection{Information retrieval}

This scenario comprises two intents: seven students used LLMs to search for concepts/materials ("Info-query"), and five students used LLMs to abstract information, primarily summarize papers ("Info-summary"). In total, 11 students provided 12 chat logs, with P13 contributing chat logs for both intents.

Regarding "Info-query", all students preferred LLMs to output answers directly when querying concepts or materials. Three students (P1/14/16) considered "getting results" to be the learning process itself. For example, using LLMs to explain concepts and implementation details (P3) or to learn cross-disciplinary topics such as data-related law (P16). Four students (P3/6/11/13) felt that the LLM's explanations (answers) could spark further thinking. For example, after seeing box plots in class, P11 asked an LLM what they represent and then asked what the interquartile range means; When LLMs explained multi-agent memory, P13 saw it could fit many scenarios and asked more questions to confirm. All students viewed LLMs in this intent as encyclopedic—like search engines, but more efficient, comprehensive, and supportive of personalized exploration.

In "Info-summary", three students (P4/5/15) used LLMs to summarize a paper's contributions, methods, and results. They did so because "LLMs can understand an entire paper in just a few seconds" (P5). Given LLMs’ extensive knowledge, they can turn identified research gaps into multiple potential ideas (P13). Therefore, two students also explore research inspiration via LLMs, from a survey (P9) or across multiple papers (P13). All students see this as asking LLMs for tailored summaries, then using those summaries to actively learn, understand, and reflect.

\section{Method - Instructors}\label{S:M-Instructors}

\subsection{Participants}
\begin{table*}[htbp]
\centering
\caption{Instructors' IDs, Gender (M = Male, F = Female), Position (Lecturer = Assistant Professor), LLM exp = experience using LLMs (1 = Unfamiliar to 5 = Expert), Teaching contents indicate the primary courses or domains taught.}
\label{tab:instructors}
\begin{tabular}{lclcl}
\toprule
ID & Gender & Position & LLM exp & Teaching contents \\
\midrule
T1 & F & Lecturer & 4 & Computer Fundamentals \\
T2 & M & Associate Professor & 4 & Computer Vision \\
T3 & F & Associate Professor & 5 & Natural Language Processing \\
T4 & M & Lecturer & 5 & Data Science \\
T5 & F & Lecturer & 3 & Introduction to Artificial Intelligence \\
T6 & M & Lecturer & 4 & Sensor Networks; Software Engineering \\
\bottomrule
\end{tabular}
\end{table*}

We recruited a total of 6 CS instructors via university email lists and social media. Instructors were asked to first complete a brief demographic survey covering gender, position, LLM usage experience (five-point Likert scale \cite{joshi_likert_2015}), and their teaching contents, as shown in Table \ref{tab:instructors}. This demographic information helped target inquiries within our study. The gender distribution was balanced (3 male, 3 female). All instructors self-reported high LLM experience (mean = 4.17). We did not collect instructors' ages because the combination of age, gender, position, and teaching content could potentially lead to identity disclosure. All instructors have teaching experience across all five scenarios. In addition, T5 emphasized that their university has a policy explicitly prohibiting students from using LLMs to complete assignments, which served as the premise for all of their insights. Each instructor received CNY 100 as compensation upon completing the session.

\subsection{User study procedure}
We conducted one-on-one sessions with each instructor, totaling 6 sessions by the time we completed the study, with each session lasting approximately 1.5 hours. Instructors could choose to attend the sessions either in person or remotely. The detailed procedure of the session and the interview questions are as follows.

\subsubsection{Intent-based interviews}
%首先，老师们被要求回忆他们发现学生在哪些场景会使用LLMs学习，以及如何使用，特别是在他们的课程中。接着我们汇总了学生数据供老师评估学生在不同场景下对待LLM生成内容的方式是否符合他们期望。我们汇总的学生数据主要包括了section \ref{S:F-Students}中的所有学生行为和观点，还包括了学生数据中归属于"Instructor or LLM"该编码的内容。但我们没有汇总并提供学生对LLMs的设计建议，因为这可能影响老师提出的设计建议存在偏见。由于大部分学生不认为LLM用于学习有很多不负责任的内容或受其伤害，我们没有向教师继续讨论该问题。此外，学生身份完全匿名化，老师不被允许查看学生的chat logs，这是因为同一所大学的老师如果曾经见过这些信息会导致学生信息泄露，但如果老师需要了解具体情况，我们会直接口述以避免将大模型回答完全告知老师.

%Building on the consolidated outcomes from the student workshops，我们已经将场景分为"Writing", "Problem-solving", "Information retrieval", "Project-based learning (PBL)", and "Coding"，包括这些场景下的各种用例。对于每个场景和用例，教师首先提出他们期望学生怎么对待LLMs的结果，然后：
%如果学生的使用方式与老师的期望产生了冲突（例如学生用LLM抄答案而老师不希望学生抄答案）：老师需要回答1）为什么不期望学生这么做？2）你一般通过哪些教育策略/方式来避免学生这种行为？3）你希望该场景下LLM应该怎么设计来更好的帮助学生学习？
%如果双方都认同从LLM直接获得答案有助于学习或者不会用LLM直接获得答案而是辅助学习场景：老师需要回答1）为什么赞同学生的行为？2）LLM在此情况下是否有更好的设计来帮助学生学习？

Firstly, instructors were asked to recall the scenarios in which they found students using LLMs and their intents, especially in the instructors' own courses. 
Next, we consolidated all text from Section \ref{S:F-Students} into a Word document and shared it with the instructors. For each intent, instructors first assessed whether student practices and views conflicted with their norms, and then answered our interview questions accordingly. However, we did not provide students' design suggestions for LLMs, because doing so could bias the instructors' own design proposals. Although most students did not believe there were many irresponsible responses in learning, we welcome instructors to present their views on irresponsible contents when articulating their norms. Additionally, student identities were anonymized, and instructors were not allowed to view chat logs to avoid re-identification; when specific context was needed, we provided brief oral descriptions instead of full LLM responses. The questions were as follows:
\begin{itemize}
    \item If student practices conflicted with instructor norms (e.g., students copying answers when instructors prefer they not copy): instructors were asked to answer 1) What are your norms, and why? 2) Which specific behaviors conflict with your norms? 3) What pedagogical strategies do you typically use to prevent such behavior? 4) How should LLMs be designed in this scenario to better support student learning?
    \item If both sides agreed that directly obtaining answers from LLMs is accepted/beneficial, or that LLM outputs should be used only as references to support learning: instructors were asked to answer 1) What are your norms, and why do you endorse student behaviors? 2) In this situation, is there better LLM design advice to help students learn?
\end{itemize}

\subsubsection{Summary and ethical discussion}
%接着老师需要总结Scenario-based interviews phase中的见解，回答：1）对于不符合期望的场景，总结一下您主要通过哪些教育方式来避免学生依赖大模型获得答案。2）分场景总结LLM应该如何设计来实现您的期望以帮助学生学习？3）在场景中，如果您认为LLM不应该直接提供答案，您认为LLM应该如何设计交互方式（例如：护栏+反问，元认知能力评估后调整策略），以更有效地鼓励思考，同时可能减少学生的抵触（因为许多学生提到可以换成直接获得答案的大模型）？同时，如果您认为不能直接提供答案，你希望LLM有类似的"妥协"机制吗，即通过一些机制可以让LLM提供答案？请说明理由。
%最后老师需要讨论伦理问题，回答：1）你怎么看待学生已经逐渐依赖LLM学习而不是向老师寻求帮助？你认为自己的职能受到威胁还是你赞成这种趋势？你认为教师相比LLM的无法被取代的优势在哪。2）你怎么评估学生-教师-大模型三者的关系？
Next, instructors were asked to synthesize insights from the intent-based interviews phase and answer: 
\begin{enumerate}
    \item For practices that do not meet norms, summarize the main pedagogical strategies you use to prevent students from relying on LLMs to obtain answers.
    \item Summarize how LLMs should be designed to realize your norms and help students learn.
    \item In intents where you believe LLMs should not provide direct answers, how should the interaction be designed to more effectively encourage thinking and prevent students from "deserting" (because many students mentioned they could switch to answer-providing LLMs)?
\end{enumerate}

Finally, instructors discussed ethical issues and answered: 
\begin{enumerate}
    \item How do you view students increasingly relying on LLMs for learning rather than seeking help from instructors? Do you feel your role is threatened, or do you support this trend? In what ways do instructors have irreplaceable advantages compared with LLMs?
    \item How do you assess the relationship among students, instructors, and LLMs?
\end{enumerate}

\subsection{Data analysis}
All sessions were recorded and transcribed. We ultimately collected 6 transcripts containing each instructor's interview data. For each session, the first author independently conducted thematic analysis \cite{braun_using_2006} of the transcript. The first coding round were largely consistent with the second coding round of the student data analysis (see Section \ref{S:DataA-Students}): we categorized instructors' statements by the same intents, with the only change that, in the first round, we added a "Summary" code to capture the instructors' concluding remarks in summary and ethical discussion phase. The second coding round differed from the student data: for each intent or "Summary", we coded statements related to our research questions and grouped them into RQ2: "Norms" and "Perspectives" and RQ4: "Pedagogical strategies" and "LLM design suggestions". We illustrate these findings in Section \ref{S:F-Instructors}.
%; RQ3: "Pedagogical strategies", "Instructor or LLM";

\section{Findings: Instructor norms}\label{S:F-Instructors}

\begin{figure*}[htbp]
  \centering
  \includegraphics[width=\linewidth]{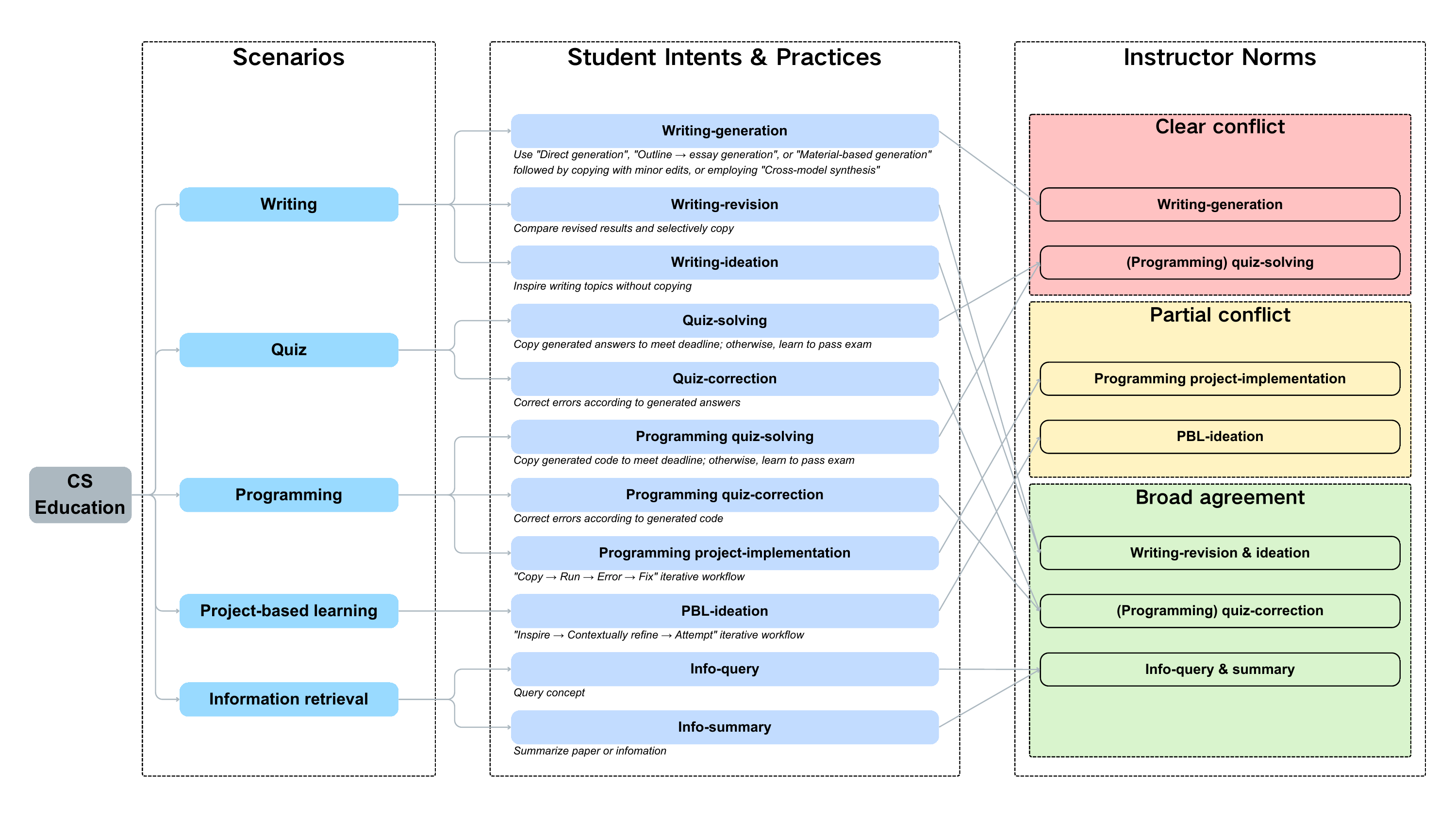}
  \caption{Scenarios, Student intents \& practices, and Instructor norms (by conflict level)}
  \Description{This is an image of a table with five columns: Conflict level, intents, Reasons for conflict, Pedagogical strategies, and LLM design recommendations. It groups intents by conflict level: Clear conflict ("Writing-generation", "(Programming) quiz-solving"); Partial conflict ("PBL-ideation", "Programming project-implementation"); and Broad agreement ("Writing-revision \& ideation", "(Programming) quiz-correction", "Info-query \& summary"). The main reason for conflict is instructors’ rejection of coping LLM outputs. The table also includes pedagogical strategies for intents with clear conflict and, for some intents, targeted LLM design suggestions.}
  \label{fig:practice-norm}
\end{figure*}

We answer RQ2 in this section by elaborating on the instructor norms corresponding to the student practices under each student intent described in Section \ref{S:F-Students}. Across all intents, we find that the core tension between student practices and instructor norms in Computer Science is: students are more "output-oriented" - prioritizing efficient information acquisition and assignment completion while overlooking some risks; whereas instructors are more "process-oriented" - emphasizing structured thinking, knowledge transfer, and justification. We group these intents into three degree of conflict between instructor norms and student practice: Clear conflict - "Writing-generation" and "(Programming) quiz-solving"; Partial conflict - "PBL-ideation" and "Programming project-implementation"; Broad agreement - "Writing-revision \& ideation", "(Programming) quiz-correction" and "Info-query \& summary". All scenarios, student intents \& practices, and instructor norms grouped by conflict level are shown in the Fig \ref{fig:practice-norm}.

\subsection{Clear conflict: "Writing-generation" and "(Programming) quiz-solving"}\label{S:CConflict}

The clear conflict arises primarily because student in these intentions relying on LLM-generated writings, answers or code to finish assignments, which contradicts pedagogical goals. Specific instructor norms are as follows.

% Still, instructors generally endorse using LLMs to check and explain solutions. This indicates a core divergence: instructors expect students to build independent competence to solve basic problems, while students prioritize completing assignments for grades with LLMs.

\subsubsection{Writing-generation}\label{S:CCWg}
As noted, most students ultimately copy LLM-generated writings with minor edits to efficiently meet assignment criteria. Five instructors (T1/2/3/5/6) strongly oppose this practice, identifying it as a prevalent dependency that bypasses the structural thinking and analysis central to writing assignments. They argued that submitting generated essays or lab reports - even if modified or synthesized from multiple models - undermines foundational capabilities. Specifically, T1 noted that "generating literature reviews impairs the ability to gather and synthesize information", while T6 observed that "reliance on LLMs for lab reports hinders independent data analysis". In addition, instructors generally support using LLMs for outlining. These views align closely with findings from previous research \cite{barrett_not_2023, malinka_educational_2023, cotton_chatting_2024, playfoot_hey_2024}.

A new finding is that instructors' opinions diverge on the "outline $\rightarrow$ essay" workflow. T1 accepts this if the outline undergoes substantial revision because "this demonstrates that students have engaged in thinking and can articulate their core ideas". Conversely, T3 and T6 value outline generation but reject the subsequent automated essay writing, noting that "this means student only mastered how to formulate writing structure." T2 and T5 remained skeptical, viewing this workflow as merely "elaborate prompt engineering" that still results in copied work. Although the majority of instructors continue to reject practices aimed at submitting copied work, a shift is evident: instructors are increasingly acknowledging the cognitive effort and thoughtfulness required to guide LLMs toward generating higher-quality output.

This shifting perspective led T4, the only instructor who accepts the submission of generated writings, to propose a distinct viewpoint: "Using prompts well to produce a high-quality essay is itself a learning skill, reflecting students' accurate understanding and the ability to obtain needed information through clear instructions". Accordingly, T4 requires students to submit the prompts and an explanation of the LLM's role in writing assignments in order to evaluate the learning process and outcomes more comprehensively. T4 elaborated:
\begin{quote}
    "If students do not understand what we ask them to write, their prompts obviously will not yield a good essay, and I will give a low grade. Conversely, if the submitted essay is excellent - even if entirely LLM-written - it shows the student learned what we intended from the assignment, at least based on my review of the prompts."
\end{quote}

\subsubsection{(Programming) quiz-solving}

\textbf{We consolidate "Quiz-solving" and "Programming quiz-solving" into a single intent "(Programming) quiz-solving"}, as students and instructors regard both center on learning through problem-solving. A clear conflict exists here, as mentioned by previous works \cite{sheard_instructor_2024, kasneci_chatgpt_2023, jost_impact_2024, bastani_generative_2025, rogers_chatgpts_2025, amoozadeh_student-ai_2024, sheese_patterns_2024}: students frequently rely on LLM-generated answers or code to meet deadlines, whereas instructors emphasize that such copying fails to train necessary reasoning and algorithmic thinking skills. Three instructors (T1/2/6) observed students copying answers for calculation, analysis, proof questions. While instructors generally endorse using LLMs to check or explain solutions after an attempt, they argue that bypassing the problem-solving process negates the educational value of assignments intended to prepare students for exams. However, we found exceptions exist for open-ended concept questions (e.g., "What are the applications and risks of LLMs in medicine?"), where instructors (T1/6) likened LLM use to a search engine and deemed it acceptable.

\subsection{Partial conflict: "Programming project-implementation" and "PBL-ideation"}\label{S:PConflict}

In these intents, which typically lack a single correct answer, instructors generally display greater tolerance, Yet there are still some conflicts. In "PBL-ideation", the conflict centers on the depth of engagement: instructors encourage using LLMs for inspiration but oppose outsourcing the core problem framing. In "Programming project-implementation", the conflict is contextual: instructors accept LLM-assisted coding for complex projects but reject it when the pedagogical goal is mastering syntax or fundamental concepts of programming.

\subsubsection{Programming project-implementation}
In contrast to the strict prohibitions in "(Programming) quiz-solving", instructors are generally more tolerant in this intent. All instructors recognized that LLMs cannot yet produce complete, complex project implementations in a single shot, making the iterative "Copy $\rightarrow$ Run $\rightarrow$ Error $\rightarrow$ Fix" iterative workflow acceptable for many. T2, T3, and T4 agreed that utilizing LLMs to implement complex code (e.g., in software, algorithms or engineering projects) aligns with professional requirements. T2 explained that "the goal of programming projects is not to test basic coding skills but overall functional and architectural design, so implementation is only one part". T3 and T4 added that successfully using tools to solve coding issues and complete projects demonstrates that students have mastered the skills needed for future jobs, thereby meeting the educational objective.

Nevertheless, conflict persists depending on specific course goals. T5 opposed copying code in experiment (e.g., implement neural networks), arguing that "copying code prevents students from learning how neural networks are implemented in code". T6 provided a clear distinction based on context: he prohibits direct copying in programming-language courses (e.g., Python and Java) where teaching student to master syntax and then coding independently, but explicitly encourages it in their software engineering course to improve efficiency. T6 elaborated on this distinction:
\begin{quote}
    "Software engineering is not about learning a programming language. It is about teaching students through the whole software development process - from requirements to final product. Coding is just one part of it."
\end{quote}

\subsubsection{PBL-ideation}
Interestingly, no instructor initially mentioned students using LLMs in the project-based learning, yet they widely accepted "Inspire $\rightarrow$ Contextually refine $\rightarrow$ Attempt" iterative workflow once discussed, viewing project-based learning as central to cultivating integrated competencies. All instructors endorsed using LLMs to choose directions and inspire ideas - leveraging the tool's breadth to generate "unexpected yet feasible project ideas" (T4) - rather than directly adopting suggestions. However, a partial conflict emerges when students transition from ideation to execution without critical oversight. Regarding a student's practice of directly adopting an LLM-suggested data structure and iteratively refining it (P8), T2, T5, and T6 acknowledged the efficiency but raised valid concerns. T2 and T5 worried that "hallucinations in proposed research plans" could lead students astray if blindly adopted. T6 further argued that "If students use LLMs to generate the technical route for an experiment when they lack ideas, that means the LLM independently performs the experiment analysis and design work, which undermines the purpose of project-based learning".

\subsection{Broad agreement: "Writing-revision \& ideation", "(Programming) quiz-correction" and "Info-query \& summary"}
Across these intents, student practices and instructor norms largely align. Both sides view LLMs as enabling tools that enhance efficiency and support self-directed learning, provided that students should maintain critical oversight. In "Writing-revision \& ideation", LLMs are accepted for translation and polishing; in "(Programming) quiz-correction", LLMs are valued for error checking after independent attempts; and in "Info-query \& summary", student practices reflect high metacognitive awareness.

\subsubsection{Writing-revision \& ideation}

Instructors acknowledged that using LLMs for translation and polishing is common practice, though they did not initially report observing students using LLMs for ideation. However, upon review, they agreed that obtaining results in both "Writing-revision" and "Writing-ideation" benefits the writing process. Therefore, \textbf{we discuss them as a intent "Writing-revision \& ideation"}. All instructors encouraged student practices - polishing, translation and ideation - in this intent, distinguishing it clearly from "Writing-generation". T2 stated that while graduate students are not expected to generate text for publication manuscripts, they are encouraged to use LLMs for polishing. Based on this, instructor emphasizes human-in-the-loop verification: students must proofread and selectively adopt changes rather than copying verbatim. T3 highlighted the necessity of this, observing that "many undergraduates use LLMs to translate but often fail to proofread carefully, leading to errors".

\subsubsection{(Programming) quiz-correction}

Same as before, \textbf{We merge "Quiz-correction" and "Programming quiz-correction" into a single intent "(Programming) quiz-correction"}, as both include the same student practices: attempting problems independently first, then utilizing LLMs to identify and fix errors. While only T3 initially mentioned this intent, all instructors endorsed it once presented. In stark contrast to norms of "(Programming) quiz-solving" where obtaining answers is discouraged, here it is valued as a feedback mechanism. T3 specifically encouraged this practice, noting that students should use LLMs to refine their own quiz solutions and learn alternative approaches.

\subsubsection{Info-query \& summary}

\textbf{We combine "Info-query" and "Info-summary" together as a intent "Info-query \& summary"}. Instructors generally viewed this intent as synonymous with self-directed learning, whether using LLMs or search engines. T2 and T5 observed students using LLMs to grasp concepts or summarize lecture slides , while T3 emphasized that "99\% of students rely on LLMs to summarize and understand paper". Instructors agreed that student practices - obtaining results in query and summary using LLMs - are beneficial, interpreting it as a sign of high metacognitive awareness where students actively recognize knowledge gaps. T2 further encouraged students to conduct literature surveys with LLMs in related research areas. However, this endorsement comes with a caveat regarding data integrity: three instructors (T1/5/6) warned that students must read critically to avoid hallucination, like incorrect explanations or fabricated bibliographic information.

%Other instructors did not initially mention students using LLMs for information retrieval. 

\section{Discussion}

\subsection{Changes in student practices and instructor norms}\label{S:Change}

Our findings reveal that the integration of LLMs into computer science education represents more than tool adoption; it is a fundamental renegotiation of student practices and instructor norms across diverse scenarios. By synthesizing existing literature with our findings, we reveal a complex dynamic (for RQ3): the collision between student pragmatism and instructor expectations has driven students to develop sophisticated strategies for LLM usage. These evolving practices, coupled with the ubiquity of LLMs, are compelling instructors to shift from initial prohibitive policies toward process-oriented adaptive strategies.

Prior research has characterized student engagement with LLMs as straightforward assistance: brainstorming, drafting, and polishing in writing \cite{barrett_not_2023}; explaining, generating, and debugging solutions for programming and quizzes \cite{bastani_generative_2025,prather_widening_2024, amoozadeh_student-ai_2024, hou_effects_2024, budhiraja_its_2024, jost_impact_2024}; and basic information retrieval \cite{youssef_examining_2024}. These practices are often framed as reliance driven by the need to pass assessments \cite{playfoot_hey_2024, stojanov_university_2024}, frequently accompanied by concealment or output modification to evade AI detection \cite{adnin_examining_2025}. While resonating with these studies, our findings further reveal that students have evolved highly complex workflows to meet academic standards and circumvent AI detection. For instance, in "Writing-generation", students employ diverse strategies such as "Direct generation", "Outline $\rightarrow$ essay generation", or "Material-based generation", and even leverage "Cross-model synthesis" to reduce AI rates. In programming, whereas existing research primarily focuses on quizzes, we identify students' "Copy $\rightarrow$ Run $\rightarrow$ Error $\rightarrow$ Fix" iterative workflow during "Programming project-implementation". Similarly, in project-based learning, students perform a concrete "Inspire $\rightarrow$ Contextually refine $\rightarrow$ Attempt" iterative workflow rather than a kind of vision \cite{zheng_charting_2024}. This indicates that students are deeply embedding LLMs into every execution phase of educational tasks.

Initially, instructor norms in response to these trends tended toward binary opposition, strictly prohibiting direct generation in writing \cite{barrett_not_2023} or coding \cite{zhou_teachers_2024, sheard_instructor_2024}, and even mandates for supervised or entirely banned AI usage \cite{harvey_dont_2025, lau_ban_2023, adnin_examining_2025}. These prohibitive policies often relied on verbal warning, AI detection \cite{adnin_examining_2025} or the introduction of LLMs with guardrail \cite{bastani_generative_2025, adnin_examining_2025, liffiton_codehelp_2024}. However, such strategies appear ineffective \cite{lau_ban_2023}; the evolving student practices to circumvent AI detection further underscore the limitations of these bans. Furthermore, our student participants explicitly indicated they would "desert" LLMs with guardrail, mirroring the challenge of ChatGPT's "study mode": students can still bypass the guardrails simply by switching back to the standard mode \cite{rogers_chatgpts_2025}. Crucially, a blanket ban risks stifling beneficial learning opportunities. As we observed, utilizing LLMs for project ideation, active information retrieval, or literature reading is widely regarded by both students and instructors as an efficient process of self-directed learning.

Given the inefficiency and side effects of enforced prohibitions, instructor norms are undergoing a fundamental shift: moving from a focus on "whether to use" to guiding "how to use for competency development" \cite{lau_ban_2023, ghimire_generative_2024}. As stated in Section \ref{S:CCWg}, while instructors continue to reject copied writing, they have begun to acknowledge the effort students invest in producing high-quality essays using LLMs. Other findings also reflect this shift: from rejecting copied code in programming quizzes to accepting the use of LLMs to complete programming projects through complex workflows; and from rejecting blind adoption to praising the use of LLMs for iteratively refining and attempting project-based learning plans. Specifically, \textbf{instructor norms are shifting from a blanket rejection of copied answers to an acceptance and appreciation of leveraging AI tools to achieve high-quality results}. In the following sections, we discuss how these changes are reshaping pedagogical strategies, LLM design (for RQ4). We also present some related changes in instructor-student relationships in Appendix \ref{A:Relation}.
%, and student-instructor relationships

\subsection{Changes in pedagogical strategies}\label{S:PedaStrategy}

%为了防止学生仅依靠复制LLM生成内容通过考核，教师与教育机构在计算机科学领域提出了一系列必要的教学策略变革：包括设计AI难以直接作答且需结合个人判断的评估任务 \cite{cotton_chatting_2024, sheard_instructor_2024}；增设代码调试任务 \cite{prather_robots_2023}；引入代码解释等口试环节 \cite{lau_ban_2023, sheard_instructor_2024}；以及提高线下考试的权重 \cite{cotton_chatting_2024, sheard_instructor_2024} 等。其中，最核心的变革在于要求学生提交作业完成过程并将其纳入评估，旨在强制披露AI使用情况，加大对盲目复制行为的惩处，从而降低过度依赖生成内容带来的风险 \cite{mahon_guidelines_2024, sheard_instructor_2024, adnin_examining_2025, simkute_new_2025}。

%正如 Section \ref{S:Change} 中教师规范的演变所强调的，老师们已逐渐承认利用AI工具产出高质量结果应被视为一项核心培养能力。基于这一共识，本研究的受访教师提出了改进后的教学策略：学生的LLM使用记录应被视为正面评估依据，而非仅用于惩罚复制行为。这意味着，展现出卓越的LLM使用技巧并产出高质量结果的学生应获得更高的评价。这一策略已被T4应用于写作课程评估中（见 Section \ref{S:CCWg}），还补充道：“我还会要求学生说明从大模型生成内容中学到了什么，以此促进反思。” 这种创新评估模式也得到了T3的认可：“若文章看似AI生成且无任何标注，我会判为0分；反之，若学生展示了出色的提示工程与工作流，则可视作加分项。” 尽管受访教师一致认为该方法不适用于基础编程测验（quiz）或知识问答，但T2、T3、T4及T6均认为其适用于编程项目评估，因为“学生提供的LLM使用记录能够反映其利用工具解决复杂任务的能力，这种能力的有效展现理应获得高分”(T4/6)。此外，鉴于目前尚未有教师提及学生在基于项目的学习场景中使用LLMs的具体情况，导致针对“PBL-ideation”意图下的学生实践的教学策略仍显空白。为此，我们呼吁教育者关注此类场景下的策略制定，以引导学生在该场景中更规范地使用LLMs。

To prevent students from passing assessments by copying LLM-generated answers, instructors and institutions in computer science have proposed a series of expected pedagogical shifts. These include replacing quizzes with tasks that are difficult for AI to answer directly and require personal judgment \cite{cotton_chatting_2024, sheard_instructor_2024}; adding code debugging tasks \cite{prather_robots_2023}; introducing oral exam such as code explanation \cite{lau_ban_2023, sheard_instructor_2024}; and increasing the weight of in-person examinations \cite{cotton_chatting_2024, sheard_instructor_2024}. Among these, a central change involves requiring students to submit their assignment completion process and incorporating it into the evaluation. This aims to mandate the disclosure of AI use, penalize blind copying, and mitigate the risks associated with excessive reliance on generated content \cite{mahon_guidelines_2024, sheard_instructor_2024, adnin_examining_2025, simkute_new_2025}.

As emphasized by the evolution of instructor norms in Section \ref{S:Change}, instructors are increasingly acknowledging that leveraging AI tools to achieve high-quality results is a core competency to be cultivated in higher education. Building on this consensus, our participating instructors proposed refined pedagogical strategies: \textbf{students' LLM usage records can be used for positive assessment rather than solely to penalize copying}. This implies that students demonstrating superior LLM proficiency alongside high-quality outcomes should be awarded higher marks. This strategy has already been implemented by T4 in their writing assessments (see Section \ref{S:CCWg}), who added, "I also require students to explain what they learned from LLMs to facilitate reflection". This innovative assessment approach was also endorsed by T3: "If an essay appears AI-generated without annotation, I assign a zero; conversely, if the student demonstrates strong prompt engineering and workflow, it is treated as a bonus". Although the instructors unanimously agreed that this approach is unsuitable for "(programming) quizzes", four of them (T2/3/4/6) argued for its applicability in "programming project" assessments, noting that "documented LLM usage reflects the student's ability of using tools for complex tasks, and the demonstration of this ability deserves high marks" (T4/6). Furthermore, as no instructors reported instances of student LLM use in project-based learning, pedagogical strategies addressing partial conflicts between student practices and instructor norms under the "PBL-ideation" intent remain absent. Consequently, we call on educators to prioritize strategy development to guide students toward more regulated LLM usage in project-based learning.

%"(Programming) quiz-solving"：考试，手写小测，或用项目取代。所有老师都提到考试是避免学生在日常直接复制LLMs输出的题解和代码的最好方式，因为考试期间是严禁使用电子产品的，学生自然无法依赖LLM解题。T2还强调手写题解，特别是代码题，可以让学生用LLMs获得答案后被迫抄写来记住解题流程。此外，T3表示自己在课堂上鼓励学生用普通的大模型和开了深度思考的大模型分别生成题目答案，她认为让学生反思LLMs生成的答案间的差异也是一种有利于学习的思考方式。T4指出考试是强制学生学会独立解决问题的最好方式，但随着LLM被广泛使用，他认为考试不应该作为一个课程的主要评分方式。T4提出了大胆的想法："我们应该积极接受学生对LLM在教育各个场景的使用，无论是获得答案还是帮助学习。而小测，简单代码这些LLM能直接完成的任务应该不被作为主要的评估手段，未来的教育应该将完成复杂项目这类LLM无法直接解决的任务作为主要评估方式。"

%All instructors noted that exams are the most effective way to prevent copying of LLM-generated solutions, since electronic devices are prohibited. T2 further emphasized that handwriting solutions - especially for programming quizzes - help them remember the logic even if students use LLMs. T4 acknowledged that exams compel independent problem-solving but argued they should not dominate grading in an era of ubiquitous LLMs. T4 proposed a bolder shift: "We should actively embrace students' use of LLMs across educational scenarios, whether to obtain answers or to support learning. Quizzes and simple coding tasks that LLMs can complete should not be major assessments; primary evaluation should move toward complex projects that LLMs cannot directly solve."

\subsection{LLM design recommendations}
%学生实践和教师规范的变化也促使他们对LLM设计提出新的建议，结合学生和教师的观点，我们总结了如下设计指南。
The evolving student practices and instructor norms have also prompted new recommendations for LLM design. Synthesizing the perspectives of both students and instructors, we summarize the following design guidelines.

\subsubsection{Intents with clear conflict: need guardrail}
%尽管我们已指出学生可以轻易绕过护栏（guardrails）迫使LLMs生成直接答案，但教师们仍主张在高等教育中部署专用LLMs，并在涉及 "Writing-generation" 和 "(Programming) quiz-solving"（即存在明确冲突的意图）时启用护栏机制。为此，我们还邀请教师参与者探讨了LLMs设计如何避免学生因受限而“弃用”（deserting）该工具，具体讨论将在下一节（section）展开。

%关于LLMs应在 "(Programming) quiz-solving" 意图下实施护栏机制，已被多项研究讨论并证实能提升学生学习表现 \cite{bastani_generative_2025, liffiton_codehelp_2024, sheese_patterns_2024,chowdhury_autotutor_2024}。此外，我们的教师参与者还提出，LLMs在 "Writing-generation" 场景下也应启用护栏——即先向学生提问，再生成大纲。据我们所知，目前尚无研究探讨过这一设计。除 T4 外，所有教师均认为LLMs不应直接生成论文。基于此，LLMs应提出针对性的、推理导向的引导（例如 "What is your central argument?", "What belongs in the first section?"），并通过多轮对话收集学生输入以草拟大纲并规划各章节，从而更好地支持构思（ideation）(6 students; T1/2/3/6)。在此过程结束后，T1/3/6 表示可以接受生成完整草稿，因为学生完成该交互过程说明学生学到了他们希望学生学到的写作技能。只有 T5 仍然不赞成学生通过此过程完成写作。

Although we have noted that students can easily bypass guardrails to get answers from LLMs, instructors still advocate for the deployment of specialized LLMs with guardrail in higher education when dealing with intents with clear conflict - "Writing-generation" and "(Programming) quiz-solving". To this end, we also invited instructor participants to share how LLM design can prevent students from "deserting" guardrail, which will be discussed in detail in the next section.

The implementation of guardrails for LLMs under the "(Programming) quiz-solving" intent has been discussed in several studies and proven to improve student learning performance \cite{bastani_generative_2025, liffiton_codehelp_2024, sheese_patterns_2024, chowdhury_autotutor_2024}. Furthermore, our instructor participants proposed that LLMs should also employ guardrails in "Writing-generation" - specifically by questioning students then generating an outline. To our knowledge, no existing research has discussed this design. Except for T4, all other instructors argued that LLMs should not generate essays directly. Therefore, LLMs should pose targeted, reasoning-driven prompts (e.g., "What is your central argument?", "What belongs in the first section?") and, through multi-turn dialogue, gather student input to draft the outline and sketch each section - thereby better supporting ideation (6 students; T1/2/3/6). After this procedure, T1/3/6 expressed acceptance of generating a full draft, as completing this interaction demonstrates that students have met the intended writing goals. Only T5 remained opposed to students completing their writing through this process.

%\textbf{"(Programming) quiz-solving": LLMs should use guardrails and question students with hints.} For CS quizzes, instructors largely converged on norms and LLM design recommendations for quiz-solving and coding items. P3 and P12 felt that receiving both answers and reasoning reduced their motivation to learn. They and all instructors favored LLMs with guardrails that use Socratic questioning to guide step-by-step learning, a practice shown to benefit learning \cite{paul_thinkers_2019, liffiton_codehelp_2024, sheese_patterns_2024, bastani_generative_2025}. For example, "the LLM states $a$, and let students explain how $a$ leads to $b$" (T1/2); In a wireless sensor network localization case, T6 suggested LLMs first ask, "How many dimensions are the object's coordinates?" LLMs can embed hints in questions (T2/3), e.g., "Are you familiar with the formula [...]?" Four instructors (T1/3/5/6) also supported revealing the final answer after the student completes the reasoning so they can check; students may also submit their own answer for correction. After the reasoning, if multiple solutions exist, the LLM should continue by asking, "Another approach is roughly [...]. Would you like to see it?" (T2/3).

\subsubsection{Prevent students from "deserting"}
%除了T5认为难以通过交互来避免学生"逃离"，其余老师提出三种避免学生逃离护栏的设计。
Except for T5, who remained skeptical about interactive solutions, other instructors proposed three design strategies to prevent students from "deserting" the guardrail.

\begin{itemize}
    \item \textbf{LLMs should default to guardrails in intents with clear conflict.} Rather than relying on students to opt into "study mode", five instructors argued that guardrails plus questioning should be the default for "Writing-generation" and "(Programming) quiz-solving". T2 further explained that a configurable "study mode" is unlikely to curb the "just want the answer" motive: "students who intend to learn will engage regardless of mode, while those who do not will simply switch to LLMs that provide answers".

    \item \textbf{LLMs should respond with empathy and make clear commitments to the learning process and outcome} to prevent "deserting" when withholding answers. T1 recommended timely encouragement and affirmation when students are stuck; T3 emphasized empathetic, collaborative language to open new perspectives (e.g., "Could we first try approaching this from [...]?") and, with a sincere tone and staged commitments, increase student willingness to engage. T4 suggested LLMs stating up front in "(Programming) quiz-solving": "I will guide your thinking instead of giving the answer as this helps you learn, and I will provide the correct solution at the end."

    \item \textbf{LLMs can add light game-like elements}, which can boost the appeal and persistence of Q\&A loop. T6 suggests morale and praise rewards when students answer follow-up questions, and setting up a simple points system: award points for correctly responding to follow-ups or completing reasoning steps, and let points unlock extra resources or feature access. This reinforces a positive cycle and helps sustain engagement.
\end{itemize}

\subsubsection{Other intents}
%对于其他用例中，老师都不反对学生直接用LLMs直接获得结果，绝大部分老师认为这些用例中LLMs无需加上护栏，并在此基础上对部分用例提出了设计建议。
For the other intents, instructors do not oppose students obtaining results from LLMs. Therefore, most consider guardrails unnecessary in these intents and, on that basis, offered design suggestions for "Info-query \& summary" and "PBL-ideation".

\textbf{"Info-query \& summary": LLMs should output answers with comprehension checks and risk warnings.} Both instructors and students (P4; T1/3/5/6) agreed that LLMs should provide answers in this intent but should follow up by prompting students to restate the information (e.g., "Can you paraphrase this concept?" or "Can you restate the paper's algorithm design?") to verify understanding. To mitigate hallucination risks, instructors emphasized including verification reminders in responses (e.g., "This is AI-generated, please verify"). Once comprehension is confirmed, LLMs can encourage deeper exploration (e.g., "Would you like to learn more about [...]?") (P3/13; T2). For concept understanding, all instructors and students (P6/11/13/14) recommended using concrete examples and analogies to avoid overly abstract explanations.

\textbf{"PBL-ideation": LLMs should output results with follow-up questions.} Five instructors and four students (P3/7/8/ 16) agreed that LLMs should generate direct answers here, though T6 emphasized preserving student autonomy in workflow design. To encourage comprehensive thinking, LLMs should prompt students to consider overlooked factors (e.g., "You also need to consider [...]") and make their internal reasoning explainable (five instructors; P7/15/16).

%此外，教师们呼吁"Writing-revision \& ideation", "(Programming) quiz-correction" and "Programming project-implementation"

\subsection{Limitations and future work}
%This research is fundamentally qualitative, 学生对LLMs输出的处理方式和观点，以及老师的期望和Norms不能代表高等教育中的所有计算机学生和老师。具体来说，我们所收集的用例仍不全面，例如我们没有收集到写作场景中学生可能用LLMs评估自己的文章，而这在投稿论文前是很常见的做法\cite{barrett_not_2023}。另外，每个地区的计算机教学方式可能有所区别，我们的学生和老师均来自中国，这使得我们的研究对中国的计算机教育提供理论基础，但其他地区需要对部分细节再次求证。
%Therefore, caution must be exercised when interpreting the generalized results. We advocate that future work can focus on 1）大规模的师生问卷调查，更全面的收集计算机教育中LLMs的使用场景和用例，汇总并制定更鲁棒的教学策略创新和LLMs设计建议。2）探索LLMs对高等教育中其他学科的影响，收集学生实践和教师norms。
This research is fundamentally qualitative. Student ways of handling LLM outputs and their views, as well as instructor norms, do not represent all CS students and instructors in higher education. Specifically, our collected intents are not exhaustive; for example, we did not capture writing intents in which students use LLMs to evaluate their own drafts, a practice that is common prior to submitting a paper \cite{barrett_not_2023}. Moreover, CS pedagogy may vary across regions. Our students and instructors are from China, so while this study offers theoretical grounding for Chinese CS education, certain details in other regions require further verification. Moreover, the instructor norms were derived after reviewing student practices, which may have introduced some bias.

Therefore, caution is warranted when interpreting the generalized results. We advocate that future work focus on 1) large-scale surveys of students and instructors to more comprehensively document LLM use scenarios and intents in CS education, and to synthesize more robust pedagogical innovations and LLM design recommendations; 2) exploring LLMs' impacts on other disciplines in higher education, and collecting corresponding student practices and instructor norms.

\section{Conclusion}

In this paper, we provide a nuanced, intent-based analysis of LLM adoption in Computer Science education, uncovering the friction between student pragmatism and instructor pedagogical goals. Our findings highlight that while broad agreement exists for using LLMs as reflective partners and information retrieval tools, conflicts persist in tasks for which students can directly obtain answers from LLMs. To adapt to students' increasingly sophisticated LLM utilization strategies, instructor norms are shifting from simple prohibitions toward recognizing students' use of LLMs to achieve high-quality outcomes, while incorporating such usage records into assessment. Furthermore, instructors propose that LLM design must adapt to current student practices: implementing default guardrails with game-like or empathetic elements to prevent students from "deserting" to answer-giving models in high-conflict intents, particularly in "Writing-generation", while integrating comprehension checks in low-conflict intents. Ultimately, this work offers the HCI and education communities an evidence-based framework for designing context-sensitive AI tools that balance efficiency with deep learning.

\begin{acks}

\end{acks}

\bibliographystyle{ACM-Reference-Format}
\bibliography{ref}

@misc{westfall_educators_2023,
    title = {Educators {Battle} {Plagiarism} {As} 89\% {Of} {Students} {Admit} {To} {Using} {OpenAI}’s {ChatGPT} {For} {Homework}},
    url = {https://www.forbes.com/sites/chriswestfall/2023/01/28/educators-battle-plagiarism-as-89-of-students-admit-to-using-open-ais-chatgpt-for-homework/},
    language = {en},
    urldate = {2025-08-31},
    journal = {Forbes},
    author = {Westfall, Chris},
    month = jan,
    year = {2023},
    note = {Section: Careers},
}

@article{yan_llm-based_2025,
    title = {{LLM}-based collaborative programming: impact on students’ computational thinking and self-efficacy},
    volume = {12},
    copyright = {2025 The Author(s)},
    issn = {2662-9992},
    shorttitle = {{LLM}-based collaborative programming},
    url = {https://www.nature.com/articles/s41599-025-04471-1},
    doi = {10.1057/s41599-025-04471-1},
    language = {en},
    number = {1},
    urldate = {2025-08-31},
    journal = {Humanities and Social Sciences Communications},
    author = {Yan, Yi-Miao and Chen, Chuang-Qi and Hu, Yang-Bang and Ye, Xin-Dong},
    month = feb,
    year = {2025},
    note = {Publisher: Palgrave},
    pages = {149},
}

@article{mogavi_chatgpt_2024,
    title = {{ChatGPT} in education: {A} blessing or a curse? {A} qualitative study exploring early adopters’ utilization and perceptions},
    volume = {2},
    issn = {29498821},
    shorttitle = {{ChatGPT} in education},
    url = {https://linkinghub.elsevier.com/retrieve/pii/S2949882123000270},
    doi = {10.1016/j.chbah.2023.100027},
    language = {en},
    number = {1},
    urldate = {2025-08-31},
    journal = {Computers in Human Behavior: Artificial Humans},
    author = {Mogavi, Reza Hadi and Deng, Chao and Juho Kim, Justin and Zhou, Pengyuan and D. Kwon, Young and Hosny Saleh Metwally, Ahmed and Tlili, Ahmed and Bassanelli, Simone and Bucchiarone, Antonio and Gujar, Sujit and Nacke, Lennart E. and Hui, Pan},
    month = jan,
    year = {2024},
    pages = {100027},
}

@article{kasneci_chatgpt_2023,
    title = {{ChatGPT} for good? {On} opportunities and challenges of large language models for education},
    volume = {103},
    issn = {10416080},
    shorttitle = {{ChatGPT} for good?},
    url = {https://linkinghub.elsevier.com/retrieve/pii/S1041608023000195},
    doi = {10.1016/j.lindif.2023.102274},
    language = {en},
    urldate = {2025-08-31},
    journal = {Learning and Individual Differences},
    author = {Kasneci, Enkelejda and Sessler, Kathrin and Küchemann, Stefan and Bannert, Maria and Dementieva, Daryna and Fischer, Frank and Gasser, Urs and Groh, Georg and Günnemann, Stephan and Hüllermeier, Eyke and Krusche, Stephan and Kutyniok, Gitta and Michaeli, Tilman and Nerdel, Claudia and Pfeffer, Jürgen and Poquet, Oleksandra and Sailer, Michael and Schmidt, Albrecht and Seidel, Tina and Stadler, Matthias and Weller, Jochen and Kuhn, Jochen and Kasneci, Gjergji},
    month = apr,
    year = {2023},
    pages = {102274},
}

@article{baig_chatgpt_2024,
    title = {{ChatGPT} in the higher education: {A} systematic literature review and research challenges},
    volume = {127},
    issn = {08830355},
    shorttitle = {{ChatGPT} in the higher education},
    url = {https://linkinghub.elsevier.com/retrieve/pii/S0883035524000971},
    doi = {10.1016/j.ijer.2024.102411},
    language = {en},
    urldate = {2025-08-31},
    journal = {International Journal of Educational Research},
    author = {Baig, Maria Ijaz and Yadegaridehkordi, Elaheh},
    year = {2024},
    pages = {102411},
}

@article{rajabi_unleashing_2024,
    title = {Unleashing {ChatGPT}'s impact in higher education: {Student} and faculty perspectives},
    volume = {2},
    issn = {29498821},
    shorttitle = {Unleashing {ChatGPT}'s impact in higher education},
    url = {https://linkinghub.elsevier.com/retrieve/pii/S2949882124000501},
    doi = {10.1016/j.chbah.2024.100090},
    language = {en},
    number = {2},
    urldate = {2025-08-31},
    journal = {Computers in Human Behavior: Artificial Humans},
    author = {Rajabi, Parsa and Taghipour, Parnian and Cukierman, Diana and Doleck, Tenzin},
    month = aug,
    year = {2024},
    pages = {100090},
}

@article{stojanov_university_2024,
    title = {University students’ self-reported reliance on {ChatGPT} for learning: {A} latent profile analysis},
    volume = {6},
    issn = {2666920X},
    shorttitle = {University students’ self-reported reliance on {ChatGPT} for learning},
    url = {https://linkinghub.elsevier.com/retrieve/pii/S2666920X24000468},
    doi = {10.1016/j.caeai.2024.100243},
    language = {en},
    urldate = {2025-08-31},
    journal = {Computers and Education: Artificial Intelligence},
    author = {Stojanov, Ana and Liu, Qian and Koh, Joyce Hwee Ling},
    month = jun,
    year = {2024},
    pages = {100243},
}

@article{youssef_examining_2024,
    title = {Examining the effect of {ChatGPT} usage on students’ academic learning and achievement: {A} survey-based study in {Ajman}, {UAE}},
    volume = {7},
    issn = {2666920X},
    shorttitle = {Examining the effect of {ChatGPT} usage on students’ academic learning and achievement},
    url = {https://linkinghub.elsevier.com/retrieve/pii/S2666920X2400119X},
    doi = {10.1016/j.caeai.2024.100316},
    language = {en},
    urldate = {2025-09-01},
    journal = {Computers and Education: Artificial Intelligence},
    author = {Youssef, Enaam and Medhat, Mervat and Abdellatif, Soumaya and Al Malek, Mahra},
    month = dec,
    year = {2024},
    pages = {100316},
}

@article{bouteraa_understanding_2024,
    title = {Understanding the diffusion of {AI}-generative ({ChatGPT}) in higher education: {Does} students' integrity matter?},
    volume = {14},
    issn = {24519588},
    shorttitle = {Understanding the diffusion of {AI}-generative ({ChatGPT}) in higher education},
    url = {https://linkinghub.elsevier.com/retrieve/pii/S2451958824000356},
    doi = {10.1016/j.chbr.2024.100402},
    language = {en},
    urldate = {2025-09-01},
    journal = {Computers in Human Behavior Reports},
    author = {Bouteraa, Mohamed and Bin-Nashwan, Saeed Awadh and Al-Daihani, Meshari and Dirie, Khadar Ahmed and Benlahcene, Abderrahim and Sadallah, Mouad and Zaki, Hafizah Omar and Lada, Suddin and Ansar, Rudy and Fook, Lim Ming and Chekima, Brahim},
    month = may,
    year = {2024},
    pages = {100402},
}

@article{cotton_chatting_2024,
    title = {Chatting and cheating: {Ensuring} academic integrity in the era of {ChatGPT}},
    volume = {61},
    issn = {1470-3297},
    shorttitle = {Chatting and cheating},
    url = {https://www.tandfonline.com/doi/full/10.1080/14703297.2023.2190148},
    doi = {10.1080/14703297.2023.2190148},
    number = {2},
    urldate = {2025-09-01},
    journal = {Innovations in Education and Teaching International},
    author = {Cotton, Debby R. E. and Cotton, Peter A. and Shipway, J. Reuben},
    month = mar,
    year = {2024},
    note = {Publisher: SRHE Website},
    pages = {228--239},
}

@inproceedings{becker_programming_2023,
    address = {Toronto ON Canada},
    title = {Programming {Is} {Hard} - {Or} at {Least} {It} {Used} to {Be}: {Educational} {Opportunities} and {Challenges} of {AI} {Code} {Generation}},
    isbn = {978-1-4503-9431-4},
    shorttitle = {Programming {Is} {Hard} - {Or} at {Least} {It} {Used} to {Be}},
    url = {https://dl.acm.org/doi/10.1145/3545945.3569759},
    doi = {10.1145/3545945.3569759},
    language = {en},
    urldate = {2025-09-01},
    booktitle = {Proceedings of the 54th {ACM} {Technical} {Symposium} on {Computer} {Science} {Education} {V}. 1},
    publisher = {ACM},
    author = {Becker, Brett A. and Denny, Paul and Finnie-Ansley, James and Luxton-Reilly, Andrew and Prather, James and Santos, Eddie Antonio},
    month = mar,
    year = {2023},
    pages = {500--506},
}

@inproceedings{lee_coauthor_2022,
    address = {New York, NY, USA},
    series = {{CHI} '22},
    title = {{CoAuthor}: {Designing} a {Human}-{AI} {Collaborative} {Writing} {Dataset} for {Exploring} {Language} {Model} {Capabilities}},
    isbn = {978-1-4503-9157-3},
    shorttitle = {{CoAuthor}},
    url = {https://dl.acm.org/doi/10.1145/3491102.3502030},
    doi = {10.1145/3491102.3502030},
    urldate = {2024-02-02},
    booktitle = {Proceedings of the 2022 {CHI} {Conference} on {Human} {Factors} in {Computing} {Systems}},
    publisher = {Association for Computing Machinery},
    author = {Lee, Mina and Liang, Percy and Yang, Qian},
    month = apr,
    year = {2022},
    pages = {1--19},
}

@misc{sultanum_datatales_2023,
    title = {{DataTales}: {Investigating} the use of {Large} {Language} {Models} for {Authoring} {Data}-{Driven} {Articles}},
    shorttitle = {{DataTales}},
    url = {http://arxiv.org/abs/2308.04076},
    doi = {10.48550/arXiv.2308.04076},
    urldate = {2024-08-25},
    publisher = {arXiv},
    author = {Sultanum, Nicole and Srinivasan, Arjun},
    month = aug,
    year = {2023},
    note = {arXiv:2308.04076 [cs]},
}

@inproceedings{dhillon_shaping_2024,
    address = {Honolulu HI USA},
    title = {Shaping {Human}-{AI} {Collaboration}: {Varied} {Scaffolding} {Levels} in {Co}-writing with {Language} {Models}},
    isbn = {979-8-4007-0330-0},
    shorttitle = {Shaping {Human}-{AI} {Collaboration}},
    url = {https://dl.acm.org/doi/10.1145/3613904.3642134},
    doi = {10.1145/3613904.3642134},
    language = {en},
    urldate = {2025-09-01},
    booktitle = {Proceedings of the {CHI} {Conference} on {Human} {Factors} in {Computing} {Systems}},
    publisher = {ACM},
    author = {Dhillon, Paramveer S. and Molaei, Somayeh and Li, Jiaqi and Golub, Maximilian and Zheng, Shaochun and Robert, Lionel Peter},
    month = may,
    year = {2024},
    pages = {1--18},
}

@inproceedings{zheng_charting_2024,
    address = {New York, NY, USA},
    series = {{CHI} '24},
    title = {Charting the {Future} of {AI} in {Project}-{Based} {Learning}: {A} {Co}-{Design} {Exploration} with {Students}},
    isbn = {979-8-4007-0330-0},
    shorttitle = {Charting the {Future} of {AI} in {Project}-{Based} {Learning}},
    url = {https://dl.acm.org/doi/10.1145/3613904.3642807},
    doi = {10.1145/3613904.3642807},
    language = {en-US},
    urldate = {2025-06-14},
    booktitle = {Proceedings of the 2024 {CHI} {Conference} on {Human} {Factors} in {Computing} {Systems}},
    publisher = {Association for Computing Machinery},
    author = {Zheng, Chengbo and Yuan, Kangyu and Guo, Bingcan and Hadi Mogavi, Reza and Peng, Zhenhui and Ma, Shuai and Ma, Xiaojuan},
    year = {2024},
    pages = {1--19},
}

@inproceedings{ravi_co-designing_2025,
    address = {New York, NY, USA},
    series = {{CHI} '25},
    title = {Co-designing {Large} {Language} {Model} {Tools} for {Project}-{Based} {Learning} with {K12} {Educators}},
    isbn = {979-8-4007-1394-1},
    url = {https://dl.acm.org/doi/10.1145/3706598.3713971},
    doi = {10.1145/3706598.3713971},
    urldate = {2025-05-27},
    booktitle = {Proceedings of the 2025 {CHI} {Conference} on {Human} {Factors} in {Computing} {Systems}},
    publisher = {Association for Computing Machinery},
    author = {Ravi, Prerna and Masla, John and Kakoti, Gisella and Lin, Grace C. and Anderson, Emma and Taylor, Matt and Ostrowski, Anastasia K. and Breazeal, Cynthia and Klopfer, Eric and Abelson, Hal},
    year = {2025},
    pages = {1--25},
}

@article{bastani_generative_2025,
    title = {Generative {AI} without guardrails can harm learning: {Evidence} from high school mathematics},
    volume = {122},
    shorttitle = {Generative {AI} without guardrails can harm learning},
    url = {https://www.pnas.org/doi/10.1073/pnas.2422633122},
    doi = {10.1073/pnas.2422633122},
    language = {en-US},
    number = {26},
    urldate = {2025-07-11},
    journal = {Proceedings of the National Academy of Sciences},
    author = {Bastani, Hamsa and Bastani, Osbert and Sungu, Alp and Ge, Haosen and Kabakcı, Özge and Mariman, Rei},
    month = jul,
    year = {2025},
    note = {Publisher: Proceedings of the National Academy of Sciences},
    pages = {e2422633122},
}

@article{playfoot_hey_2024,
    title = {Hey {ChatGPT}, give me a title for a paper about degree apathy and student use of {AI} for assignment writing},
    volume = {62},
    issn = {10967516},
    url = {https://linkinghub.elsevier.com/retrieve/pii/S1096751624000125},
    doi = {10.1016/j.iheduc.2024.100950},
    language = {en},
    urldate = {2025-09-01},
    journal = {The Internet and Higher Education},
    author = {Playfoot, David and Quigley, Martyn and Thomas, Andrew G.},
    month = jun,
    year = {2024},
    pages = {100950},
}

@inproceedings{malinka_educational_2023,
    address = {New York, NY, USA},
    series = {{ITiCSE} 2023},
    title = {On the {Educational} {Impact} of {ChatGPT}: {Is} {Artificial} {Intelligence} {Ready} to {Obtain} a {University} {Degree}?},
    isbn = {979-8-4007-0138-2},
    shorttitle = {On the {Educational} {Impact} of {ChatGPT}},
    url = {https://dl.acm.org/doi/10.1145/3587102.3588827},
    doi = {10.1145/3587102.3588827},
    urldate = {2025-08-31},
    booktitle = {Proceedings of the 2023 {Conference} on {Innovation} and {Technology} in {Computer} {Science} {Education} {V}. 1},
    publisher = {Association for Computing Machinery},
    author = {Malinka, Kamil and Peresíni, Martin and Firc, Anton and Hujnák, Ondrej and Janus, Filip},
    year = {2023},
    pages = {47--53},
}

@article{deng_does_2025,
    title = {Does {ChatGPT} enhance student learning? {A} systematic review and meta-analysis of experimental studies},
    volume = {227},
    issn = {03601315},
    shorttitle = {Does {ChatGPT} enhance student learning?},
    url = {https://linkinghub.elsevier.com/retrieve/pii/S0360131524002380},
    doi = {10.1016/j.compedu.2024.105224},
    language = {en},
    urldate = {2025-09-01},
    journal = {Computers \& Education},
    author = {Deng, Ruiqi and Jiang, Maoli and Yu, Xinlu and Lu, Yuyan and Liu, Shasha},
    month = apr,
    year = {2025},
    pages = {105224},
}

@misc{kosmyna_your_2025,
    title = {Your {Brain} on {ChatGPT}: {Accumulation} of {Cognitive} {Debt} when {Using} an {AI} {Assistant} for {Essay} {Writing} {Task}},
    shorttitle = {Your {Brain} on {ChatGPT}},
    url = {http://arxiv.org/abs/2506.08872},
    doi = {10.48550/arXiv.2506.08872},
    urldate = {2025-06-22},
    publisher = {arXiv},
    author = {Kosmyna, Nataliya and Hauptmann, Eugene and Yuan, Ye Tong and Situ, Jessica and Liao, Xian-Hao and Beresnitzky, Ashly Vivian and Braunstein, Iris and Maes, Pattie},
    month = jun,
    year = {2025},
    note = {arXiv:2506.08872 [cs]},
}

@inproceedings{cao_ai_2025,
    address = {New York, NY, USA},
    series = {{CHI} '25},
    title = {{AI} {Literacy} for {Underserved} {Students}: {Leveraging} {Cultural} {Capital} from {Underserved} {Communities} for {AI} {Education} {Research}},
    isbn = {979-8-4007-1394-1},
    shorttitle = {{AI} {Literacy} for {Underserved} {Students}},
    url = {https://dl.acm.org/doi/10.1145/3706598.3713173},
    doi = {10.1145/3706598.3713173},
    language = {en-US},
    urldate = {2025-06-14},
    booktitle = {Proceedings of the 2025 {CHI} {Conference} on {Human} {Factors} in {Computing} {Systems}},
    publisher = {Association for Computing Machinery},
    author = {Cao, Huajie Jay and Choi, Kahyun and Park, Claire and Lee, Hee Rin},
    year = {2025},
    pages = {1--15},
}

@inproceedings{chowdhury_autotutor_2024,
    address = {New York, NY, USA},
    series = {L@{S} '24},
    title = {{AutoTutor} meets {Large} {Language} {Models}: {A} {Language} {Model} {Tutor} with {Rich} {Pedagogy} and {Guardrails}},
    isbn = {979-8-4007-0633-2},
    shorttitle = {{AutoTutor} meets {Large} {Language} {Models}},
    url = {https://dl.acm.org/doi/10.1145/3657604.3662041},
    doi = {10.1145/3657604.3662041},
    urldate = {2025-08-31},
    booktitle = {Proceedings of the {Eleventh} {ACM} {Conference} on {Learning} @ {Scale}},
    publisher = {Association for Computing Machinery},
    author = {Chowdhury, Sankalan Pal and Zouhar, Vilém and Sachan, Mrinmaya},
    year = {2024},
    pages = {5--15},
}

@article{kumar_guiding_2024,
    title = {Guiding {Students} in {Using} {LLMs} in {Supported} {Learning} {Environments}: {Effects} on {Interaction} {Dynamics}, {Learner} {Performance}, {Confidence}, and {Trust}},
    volume = {8},
    shorttitle = {Guiding {Students} in {Using} {LLMs} in {Supported} {Learning} {Environments}},
    url = {https://dl.acm.org/doi/10.1145/3687038},
    doi = {10.1145/3687038},
    number = {CSCW2},
    urldate = {2025-06-13},
    journal = {Proc. ACM Hum.-Comput. Interact.},
    author = {Kumar, Harsh and Musabirov, Ilya and Reza, Mohi and Shi, Jiakai and Wang, Xinyuan and Williams, Joseph Jay and Kuzminykh, Anastasia and Liut, Michael},
    year = {2024},
    pages = {499:1--499:30},
}

@inproceedings{kumar_supporting_2024,
    address = {New York, NY, USA},
    series = {L@{S} '24},
    title = {Supporting {Self}-{Reflection} at {Scale} with {Large} {Language} {Models}: {Insights} from {Randomized} {Field} {Experiments} in {Classrooms}},
    isbn = {979-8-4007-0633-2},
    shorttitle = {Supporting {Self}-{Reflection} at {Scale} with {Large} {Language} {Models}},
    url = {https://dl.acm.org/doi/10.1145/3657604.3662042},
    doi = {10.1145/3657604.3662042},
    language = {en-US},
    urldate = {2025-06-14},
    booktitle = {Proceedings of the {Eleventh} {ACM} {Conference} on {Learning} @ {Scale}},
    publisher = {Association for Computing Machinery},
    author = {Kumar, Harsh and Xiao, Ruiwei and Lawson, Benjamin and Musabirov, Ilya and Shi, Jiakai and Wang, Xinyuan and Luo, Huayin and Williams, Joseph Jay and Rafferty, Anna N. and Stamper, John and Liut, Michael},
    year = {2024},
    pages = {86--97},
}

@inproceedings{prabhudesai_here_2025,
    address = {New York, NY, USA},
    series = {{CHI} '25},
    title = {"{Here} the {GPT} made a choice, and every choice can be biased": {How} {Students} {Critically} {Engage} with {LLMs} through {End}-{User} {Auditing} {Activity}},
    isbn = {979-8-4007-1394-1},
    shorttitle = {"{Here} the {GPT} made a choice, and every choice can be biased"},
    url = {https://dl.acm.org/doi/10.1145/3706598.3713714},
    doi = {10.1145/3706598.3713714},
    language = {en-US},
    urldate = {2025-06-12},
    booktitle = {Proceedings of the 2025 {CHI} {Conference} on {Human} {Factors} in {Computing} {Systems}},
    publisher = {Association for Computing Machinery},
    author = {Prabhudesai, Snehal and Kasi, Ananya Prashant and Mansingh, Anmol and Das Antar, Anindya and Shen, Hua and Banovic, Nikola},
    year = {2025},
    pages = {1--23},
}

@inproceedings{chaudhury_milestones_2025,
    address = {New York, NY, USA},
    series = {{CHI} '25},
    title = {{MILESTONES}: {The} {Design} and {Field} {Evaluation} of a {Semi}-{Automated} {Tool} for {Promoting} {Self}-{Directed} {Learning} {Among} {Online} {Learners}},
    isbn = {979-8-4007-1394-1},
    shorttitle = {{MILESTONES}},
    url = {https://dl.acm.org/doi/10.1145/3706598.3714295},
    doi = {10.1145/3706598.3714295},
    urldate = {2025-06-13},
    booktitle = {Proceedings of the 2025 {CHI} {Conference} on {Human} {Factors} in {Computing} {Systems}},
    publisher = {Association for Computing Machinery},
    author = {Chaudhury, Rimika and Huffman, Courtenay and Kwan, Isabelle and Deol, Gurnoor S. and Dhillon, Supreet and Chilana, Parmit K},
    year = {2025},
    pages = {1--16},
}

@inproceedings{park_promise_2024,
    address = {New York, NY, USA},
    series = {{CHI} '24},
    title = {The {Promise} and {Peril} of {ChatGPT} in {Higher} {Education}: {Opportunities}, {Challenges}, and {Design} {Implications}},
    isbn = {979-8-4007-0330-0},
    shorttitle = {The {Promise} and {Peril} of {ChatGPT} in {Higher} {Education}},
    url = {https://dl.acm.org/doi/10.1145/3613904.3642785},
    doi = {10.1145/3613904.3642785},
    urldate = {2025-01-13},
    booktitle = {Proceedings of the 2024 {CHI} {Conference} on {Human} {Factors} in {Computing} {Systems}},
    publisher = {Association for Computing Machinery},
    author = {Park, Hyanghee and Ahn, Daehwan},
    year = {2024},
    pages = {1--21},
}

@article{borge_using_2024,
    title = {Using generative ai as a simulation to support higher-order thinking},
    volume = {19},
    copyright = {2024 International Society of the Learning Sciences, Inc.},
    issn = {1556-1615},
    url = {https://link.springer.com/article/10.1007/s11412-024-09437-0},
    doi = {10.1007/s11412-024-09437-0},
    language = {en},
    number = {4},
    urldate = {2025-07-29},
    journal = {International Journal of Computer-Supported Collaborative Learning},
    author = {Borge, M. and Smith, B. K. and Aldemir, T.},
    month = dec,
    year = {2024},
    note = {Company: Springer
Distributor: Springer
Institution: Springer
Label: Springer
Number: 4
Publisher: Springer US},
    pages = {479--532},
}

@article{neshaei_metacognition_2025,
    title = {Metacognition meets {AI}: {Empowering} reflective writing with large language models},
    volume = {56},
    copyright = {© 2025 British Educational Research Association.},
    issn = {1467-8535},
    shorttitle = {Metacognition meets {AI}},
    url = {https://onlinelibrary.wiley.com/doi/abs/10.1111/bjet.13601},
    doi = {10.1111/bjet.13601},
    language = {en},
    number = {5},
    urldate = {2025-06-23},
    journal = {British Journal of Educational Technology},
    author = {Neshaei, Seyed Parsa and Mejia-Domenzain, Paola and Davis, Richard Lee and Käser, Tanja},
    month = may,
    year = {2025},
    note = {\_eprint: https://bera-journals.onlinelibrary.wiley.com/doi/pdf/10.1111/bjet.13601},
    pages = {1864--1896},
}

@inproceedings{neshaei_mindmate_2025,
    address = {New York, NY, USA},
    series = {{CHI} {EA} '25},
    title = {{MindMate}: {Exploring} the {Effect} of {Conversational} {Agents} on {Reflective} {Writing}},
    isbn = {979-8-4007-1395-8},
    shorttitle = {{MindMate}},
    url = {https://dl.acm.org/doi/10.1145/3706599.3720029},
    doi = {10.1145/3706599.3720029},
    urldate = {2025-06-23},
    booktitle = {Proceedings of the {Extended} {Abstracts} of the {CHI} {Conference} on {Human} {Factors} in {Computing} {Systems}},
    publisher = {Association for Computing Machinery},
    author = {Neshaei, Seyed Parsa and Wambsganss, Thiemo and El Bouchrifi, Hind and Käser, Tanja},
    year = {2025},
    pages = {1--9},
}

@article{kestin_ai_2025,
    title = {{AI} tutoring outperforms in-class active learning: an {RCT} introducing a novel research-based design in an authentic educational setting},
    volume = {15},
    copyright = {2025 The Author(s)},
    issn = {2045-2322},
    shorttitle = {{AI} tutoring outperforms in-class active learning},
    url = {https://www.nature.com/articles/s41598-025-97652-6},
    doi = {10.1038/s41598-025-97652-6},
    language = {en},
    number = {1},
    urldate = {2025-07-10},
    journal = {Scientific Reports},
    author = {Kestin, Greg and Miller, Kelly and Klales, Anna and Milbourne, Timothy and Ponti, Gregorio},
    month = jun,
    year = {2025},
    note = {Publisher: Nature Publishing Group},
    pages = {17458},
}

@book{boud_reflection_2013,
    address = {Hoboken},
    title = {Reflection: {Turning} {Experience} into {Learning}},
    isbn = {978-0-85038-864-0},
    shorttitle = {Reflection},
    language = {en},
    publisher = {Taylor and Francis},
    author = {Boud, David and Keogh, Rosemary and Walker, David},
    year = {2013},
}

@article{gregory_assessing_1994,
    title = {Assessing {Metacognitive} {Awareness}},
    volume = {19},
    issn = {0361-476X},
    url = {https://www.sciencedirect.com/science/article/abs/pii/S0361476X84710332},
    doi = {10.1006/ceps.1994.1033},
    language = {en-US},
    number = {4},
    urldate = {2025-06-23},
    journal = {Contemporary Educational Psychology},
    author = {Gregory, Schraw and Sperling, Dennison Rayne},
    month = oct,
    year = {1994},
    note = {Publisher: Academic Press},
    pages = {460--475},
}

@misc{rogers_chatgpts_2025,
    title = {{ChatGPT}’s {Study} {Mode} {Is} {Here}. {It} {Won}’t {Fix} {Education}’s {AI} {Problems}},
    url = {https://www.wired.com/story/chatgpt-study-mode/},
    language = {en-US},
    urldate = {2025-09-01},
    journal = {Wired},
    author = {Rogers, Reece},
    month = jul,
    year = {2025},
    note = {Section: tags},
}

@book{dunlosky_metacognition_2008,
    address = {CA},
    title = {Metacognition},
    isbn = {978-1-4833-6004-1},
    language = {en},
    publisher = {SAGE Publications},
    author = {Dunlosky, John and Metcalfe, Janet},
    month = sep,
    year = {2008},
    note = {Google-Books-ID: eVUXBAAAQBAJ},
}

@inproceedings{prather_robots_2023,
    address = {New York, NY, USA},
    series = {{ITiCSE}-{WGR} '23},
    title = {The {Robots} {Are} {Here}: {Navigating} the {Generative} {AI} {Revolution} in {Computing} {Education}},
    isbn = {979-8-4007-0405-5},
    shorttitle = {The {Robots} {Are} {Here}},
    url = {https://dl.acm.org/doi/10.1145/3623762.3633499},
    doi = {10.1145/3623762.3633499},
    urldate = {2025-09-01},
    booktitle = {Proceedings of the 2023 {Working} {Group} {Reports} on {Innovation} and {Technology} in {Computer} {Science} {Education}},
    publisher = {Association for Computing Machinery},
    author = {Prather, James and Denny, Paul and Leinonen, Juho and Becker, Brett A. and Albluwi, Ibrahim and Craig, Michelle and Keuning, Hieke and Kiesler, Natalie and Kohn, Tobias and Luxton-Reilly, Andrew and MacNeil, Stephen and Petersen, Andrew and Pettit, Raymond and Reeves, Brent N. and Savelka, Jaromir},
    year = {2023},
    pages = {108--159},
}

@inproceedings{stone_exploring_2024,
    address = {New York, NY, USA},
    series = {{UKICER} '24},
    title = {Exploring {Human}-{Centered} {Approaches} in {Generative} {AI} and {Introductory} {Programming} {Research}: {A} {Scoping} {Review}},
    isbn = {979-8-4007-1177-0},
    shorttitle = {Exploring {Human}-{Centered} {Approaches} in {Generative} {AI} and {Introductory} {Programming} {Research}},
    url = {https://dl.acm.org/doi/10.1145/3689535.3689553},
    doi = {10.1145/3689535.3689553},
    urldate = {2025-09-01},
    booktitle = {Proceedings of the 2024 {Conference} on {United} {Kingdom} \& {Ireland} {Computing} {Education} {Research}},
    publisher = {Association for Computing Machinery},
    author = {Stone, Irene},
    year = {2024},
    pages = {1--7},
}

@inproceedings{budhiraja_its_2024,
    address = {New York, NY, USA},
    series = {{ACE} '24},
    title = {“{It}'s not like {Jarvis}, but it's pretty close!” - {Examining} {ChatGPT}'s {Usage} among {Undergraduate} {Students} in {Computer} {Science}},
    isbn = {979-8-4007-1619-5},
    url = {https://dl.acm.org/doi/10.1145/3636243.3636257},
    doi = {10.1145/3636243.3636257},
    urldate = {2025-09-01},
    booktitle = {Proceedings of the 26th {Australasian} {Computing} {Education} {Conference}},
    publisher = {Association for Computing Machinery},
    author = {Budhiraja, Ritvik and Joshi, Ishika and Challa, Jagat Sesh and Akolekar, Harshal D. and Kumar, Dhruv},
    year = {2024},
    pages = {124--133},
}

@misc{simkute_new_2025,
    title = {The {New} {Calculator}? {Practices}, {Norms}, and {Implications} of {Generative} {AI} in {Higher} {Education}},
    shorttitle = {The {New} {Calculator}?},
    url = {http://arxiv.org/abs/2501.08864},
    doi = {10.48550/arXiv.2501.08864},
    urldate = {2025-09-01},
    publisher = {arXiv},
    author = {Simkute, Auste and Kewenig, Viktor and Sellen, Abigail and Rintel, Sean and Tankelevitch, Lev},
    month = jan,
    year = {2025},
    note = {arXiv:2501.08864 [cs]},
}

@article{stadler_cognitive_2024,
    title = {Cognitive ease at a cost: {LLMs} reduce mental effort but compromise depth in student scientific inquiry},
    volume = {160},
    issn = {0747-5632},
    shorttitle = {Cognitive ease at a cost},
    url = {https://www.sciencedirect.com/science/article/pii/S0747563224002541},
    doi = {10.1016/j.chb.2024.108386},
    language = {en-US},
    urldate = {2025-06-12},
    journal = {Computers in Human Behavior},
    author = {Stadler, Matthias and Bannert, Maria and Sailer, Michael},
    month = nov,
    year = {2024},
    note = {Publisher: Pergamon},
    pages = {108386},
}

@article{adel_chatgpt_2024,
    title = {{ChatGPT} {Promises} and {Challenges} in {Education}: {Computational} and {Ethical} {Perspectives}},
    volume = {14},
    copyright = {http://creativecommons.org/licenses/by/3.0/},
    issn = {2227-7102},
    shorttitle = {{ChatGPT} {Promises} and {Challenges} in {Education}},
    url = {https://www.mdpi.com/2227-7102/14/8/814},
    doi = {10.3390/educsci14080814},
    language = {en},
    number = {8},
    urldate = {2025-09-02},
    journal = {Education Sciences},
    author = {Adel, Amr and Ahsan, Ali and Davison, Claire},
    month = aug,
    year = {2024},
    note = {Publisher: Multidisciplinary Digital Publishing Institute},
    pages = {814},
}

@inproceedings{adnin_examining_2025,
    address = {New York, NY, USA},
    series = {{CHI} '25},
    title = {Examining {Student} and {Teacher} {Perspectives} on {Undisclosed} {Use} of {Generative} {AI} in {Academic} {Work}},
    isbn = {979-8-4007-1394-1},
    url = {https://dl.acm.org/doi/10.1145/3706598.3713393},
    doi = {10.1145/3706598.3713393},
    urldate = {2025-09-02},
    booktitle = {Proceedings of the 2025 {CHI} {Conference} on {Human} {Factors} in {Computing} {Systems}},
    publisher = {Association for Computing Machinery},
    author = {Adnin, Rudaiba and Pandkar, Atharva and Yao, Bingsheng and Wang, Dakuo and Das, Maitraye},
    year = {2025},
    pages = {1--17},
}

@article{wang_effect_2025,
    title = {The effect of {ChatGPT} on students’ learning performance, learning perception, and higher-order thinking: insights from a meta-analysis},
    volume = {12},
    issn = {2662-9992},
    shorttitle = {The effect of {ChatGPT} on students’ learning performance, learning perception, and higher-order thinking},
    url = {https://www.nature.com/articles/s41599-025-04787-y},
    doi = {10.1057/s41599-025-04787-y},
    language = {en},
    number = {1},
    urldate = {2025-09-02},
    journal = {Humanities and Social Sciences Communications},
    author = {Wang, Jin and Fan, Wenxiang},
    month = may,
    year = {2025},
    pages = {621},
}

@inproceedings{liffiton_codehelp_2024,
    address = {New York, NY, USA},
    series = {Koli {Calling} '23},
    title = {{CodeHelp}: {Using} {Large} {Language} {Models} with {Guardrails} for {Scalable} {Support} in {Programming} {Classes}},
    isbn = {979-8-4007-1653-9},
    shorttitle = {{CodeHelp}},
    url = {https://dl.acm.org/doi/10.1145/3631802.3631830},
    doi = {10.1145/3631802.3631830},
    urldate = {2025-09-02},
    booktitle = {Proceedings of the 23rd {Koli} {Calling} {International} {Conference} on {Computing} {Education} {Research}},
    publisher = {Association for Computing Machinery},
    author = {Liffiton, Mark and Sheese, Brad E and Savelka, Jaromir and Denny, Paul},
    year = {2024},
    pages = {1--11},
}

@article{chan_will_2024,
    title = {Will generative {AI} replace teachers in higher education? {A} study of teacher and student perceptions},
    volume = {83},
    issn = {0191491X},
    shorttitle = {Will generative {AI} replace teachers in higher education?},
    url = {https://linkinghub.elsevier.com/retrieve/pii/S0191491X24000749},
    doi = {10.1016/j.stueduc.2024.101395},
    language = {en},
    urldate = {2025-09-02},
    journal = {Studies in Educational Evaluation},
    author = {Chan, Cecilia Ka Yuk and Tsi, Louisa H.Y.},
    month = dec,
    year = {2024},
    pages = {101395},
}

@article{barrett_not_2023,
    title = {Not quite eye to {A}.{I}.: student and teacher perspectives on the use of generative artificial intelligence in the writing process},
    volume = {20},
    issn = {2365-9440},
    shorttitle = {Not quite eye to {A}.{I}.},
    url = {https://doi.org/10.1186/s41239-023-00427-0},
    doi = {10.1186/s41239-023-00427-0},
    number = {1},
    urldate = {2025-09-02},
    journal = {International Journal of Educational Technology in Higher Education},
    author = {Barrett, Alex and Pack, Austin},
    month = nov,
    year = {2023},
    pages = {59},
}

@inproceedings{hou_effects_2024,
    address = {New York, NY, USA},
    series = {{ACE} '24},
    title = {The {Effects} of {Generative} {AI} on {Computing} {Students}’ {Help}-{Seeking} {Preferences}},
    isbn = {979-8-4007-1619-5},
    url = {https://dl.acm.org/doi/10.1145/3636243.3636248},
    doi = {10.1145/3636243.3636248},
    urldate = {2025-09-02},
    booktitle = {Proceedings of the 26th {Australasian} {Computing} {Education} {Conference}},
    publisher = {Association for Computing Machinery},
    author = {Hou, Irene and Mettille, Sophia and Man, Owen and Li, Zhuo and Zastudil, Cynthia and MacNeil, Stephen},
    year = {2024},
    pages = {39--48},
}

@misc{chan_students_2023,
    title = {Students' {Voices} on {Generative} {AI}: {Perceptions}, {Benefits}, and {Challenges} in {Higher} {Education}},
    shorttitle = {Students' {Voices} on {Generative} {AI}},
    url = {http://arxiv.org/abs/2305.00290},
    doi = {10.48550/arXiv.2305.00290},
    urldate = {2025-09-02},
    publisher = {arXiv},
    author = {Chan, Cecilia Ka Yuk and Hu, Wenjie},
    month = apr,
    year = {2023},
    note = {arXiv:2305.00290 [cs]},
}

@inproceedings{kubullek_understanding_2024,
    address = {New York, NY, USA},
    series = {{MuC} '24},
    title = {Understanding the {Adoption} of {ChatGPT} in {Higher} {Education}: {A} {Comparative} {Study} with {Insights} from {STEM} and {Business} {Students}},
    isbn = {979-8-4007-0998-2},
    shorttitle = {Understanding the {Adoption} of {ChatGPT} in {Higher} {Education}},
    url = {https://dl.acm.org/doi/10.1145/3670653.3677507},
    doi = {10.1145/3670653.3677507},
    urldate = {2025-09-02},
    booktitle = {Proceedings of {Mensch} und {Computer} 2024},
    publisher = {Association for Computing Machinery},
    author = {Kubullek, Ann-Kathrin and Kumaç, Nadire and Dogangün, Aysegül},
    year = {2024},
    pages = {684--689},
}

@inproceedings{amoozadeh_trust_2024,
    address = {New York, NY, USA},
    series = {{SIGCSE} 2024},
    title = {Trust in {Generative} {AI} among {Students}: {An} exploratory study},
    isbn = {979-8-4007-0423-9},
    shorttitle = {Trust in {Generative} {AI} among {Students}},
    url = {https://dl.acm.org/doi/10.1145/3626252.3630842},
    doi = {10.1145/3626252.3630842},
    urldate = {2025-09-02},
    booktitle = {Proceedings of the 55th {ACM} {Technical} {Symposium} on {Computer} {Science} {Education} {V}. 1},
    publisher = {Association for Computing Machinery},
    author = {Amoozadeh, Matin and Daniels, David and Nam, Daye and Kumar, Aayush and Chen, Stella and Hilton, Michael and Srinivasa Ragavan, Sruti and Alipour, Mohammad Amin},
    year = {2024},
    pages = {67--73},
}

@article{johnston_student_2024,
    title = {Student perspectives on the use of generative artificial intelligence technologies in higher education},
    volume = {20},
    copyright = {2024 The Author(s)},
    issn = {1833-2595},
    url = {https://edintegrity.biomedcentral.com/articles/10.1007/s40979-024-00149-4},
    doi = {10.1007/s40979-024-00149-4},
    language = {en},
    number = {1},
    urldate = {2025-09-02},
    journal = {International Journal for Educational Integrity},
    author = {Johnston, Heather and Wells, Rebecca F. and Shanks, Elizabeth M. and Boey, Timothy and Parsons, Bryony N.},
    month = dec,
    year = {2024},
    note = {Publisher: BioMed Central},
    pages = {1--21},
}

@article{wu_systematic_2025,
    title = {A {Systematic} {Review} of {Responses}, {Attitudes}, and {Utilization} {Behaviors} on {Generative} {AI} for {Teaching} and {Learning} in {Higher} {Education}},
    volume = {15},
    issn = {2076-328X},
    url = {https://www.ncbi.nlm.nih.gov/pmc/articles/PMC12023922/},
    doi = {10.3390/bs15040467},
    number = {4},
    urldate = {2025-06-19},
    journal = {Behavioral Sciences},
    author = {Wu, Fan and Dang, Yang and Li, Manli},
    month = apr,
    year = {2025},
    pmid = {40282088},
    pmcid = {PMC12023922},
    pages = {467},
}

@inproceedings{petrovska_incorporating_2024,
    address = {New York, NY, USA},
    series = {{CEP} '24},
    title = {Incorporating {Generative} {AI} into {Software} {Development} {Education}},
    isbn = {979-8-4007-0932-6},
    url = {https://dl.acm.org/doi/10.1145/3633053.3633057},
    doi = {10.1145/3633053.3633057},
    urldate = {2025-09-02},
    booktitle = {Proceedings of the 8th {Conference} on {Computing} {Education} {Practice}},
    publisher = {Association for Computing Machinery},
    author = {Petrovska, Olga and Clift, Lee and Moller, Faron and Pearsall, Rebecca},
    year = {2024},
    pages = {37--40},
}

@inproceedings{harvey_dont_2025,
    address = {New York, NY, USA},
    series = {{CHI} '25},
    title = {"{Don}'t {Forget} the {Teachers}": {Towards} an {Educator}-{Centered} {Understanding} of {Harms} from {Large} {Language} {Models} in {Education}},
    isbn = {979-8-4007-1394-1},
    shorttitle = {"{Don}'t {Forget} the {Teachers}"},
    url = {https://dl.acm.org/doi/10.1145/3706598.3713210},
    doi = {10.1145/3706598.3713210},
    urldate = {2025-09-02},
    booktitle = {Proceedings of the 2025 {CHI} {Conference} on {Human} {Factors} in {Computing} {Systems}},
    publisher = {Association for Computing Machinery},
    author = {Harvey, Emma and Koenecke, Allison and Kizilcec, Rene F.},
    year = {2025},
    pages = {1--19},
}

@article{hasanein_drivers_2023,
    title = {Drivers and {Consequences} of {ChatGPT} {Use} in {Higher} {Education}: {Key} {Stakeholder} {Perspectives}},
    volume = {13},
    copyright = {http://creativecommons.org/licenses/by/3.0/},
    issn = {2254-9625},
    shorttitle = {Drivers and {Consequences} of {ChatGPT} {Use} in {Higher} {Education}},
    url = {https://www.mdpi.com/2254-9625/13/11/181},
    doi = {10.3390/ejihpe13110181},
    language = {en},
    number = {11},
    urldate = {2025-09-02},
    journal = {European Journal of Investigation in Health, Psychology and Education},
    author = {Hasanein, Ahmed M. and Sobaih, Abu Elnasr E.},
    month = nov,
    year = {2023},
    note = {Publisher: Multidisciplinary Digital Publishing Institute},
    pages = {2599--2614},
}

@inproceedings{sheese_patterns_2024,
    address = {New York, NY, USA},
    series = {{ACE} '24},
    title = {Patterns of {Student} {Help}-{Seeking} {When} {Using} a {Large} {Language} {Model}-{Powered} {Programming} {Assistant}},
    isbn = {979-8-4007-1619-5},
    url = {https://dl.acm.org/doi/10.1145/3636243.3636249},
    doi = {10.1145/3636243.3636249},
    urldate = {2025-09-02},
    booktitle = {Proceedings of the 26th {Australasian} {Computing} {Education} {Conference}},
    publisher = {Association for Computing Machinery},
    author = {Sheese, Brad and Liffiton, Mark and Savelka, Jaromir and Denny, Paul},
    year = {2024},
    pages = {49--57},
}

@article{singh_exploring_2023,
    title = {Exploring {Computer} {Science} {Students}’ {Perception} of {ChatGPT} in {Higher} {Education}: {A} {Descriptive} and {Correlation} {Study}},
    volume = {13},
    copyright = {http://creativecommons.org/licenses/by/3.0/},
    issn = {2227-7102},
    shorttitle = {Exploring {Computer} {Science} {Students}’ {Perception} of {ChatGPT} in {Higher} {Education}},
    url = {https://www.mdpi.com/2227-7102/13/9/924},
    doi = {10.3390/educsci13090924},
    language = {en},
    number = {9},
    urldate = {2025-09-04},
    journal = {Education Sciences},
    author = {Singh, Harpreet and Tayarani-Najaran, Mohammad-Hassan and Yaqoob, Muhammad},
    month = sep,
    year = {2023},
    note = {Publisher: Multidisciplinary Digital Publishing Institute},
    pages = {924},
}

@inproceedings{sheard_instructor_2024,
    address = {New York, NY, USA},
    series = {{SIGCSE} 2024},
    title = {Instructor {Perceptions} of {AI} {Code} {Generation} {Tools} - {A} {Multi}-{Institutional} {Interview} {Study}},
    isbn = {979-8-4007-0423-9},
    url = {https://dl.acm.org/doi/10.1145/3626252.3630880},
    doi = {10.1145/3626252.3630880},
    urldate = {2025-09-03},
    booktitle = {Proceedings of the 55th {ACM} {Technical} {Symposium} on {Computer} {Science} {Education} {V}. 1},
    publisher = {Association for Computing Machinery},
    author = {Sheard, Judy and Denny, Paul and Hellas, Arto and Leinonen, Juho and Malmi, Lauri and Simon},
    year = {2024},
    pages = {1223--1229},
}

@inproceedings{mahon_guidelines_2024,
    address = {New York, NY, USA},
    series = {{ITiCSE} 2024},
    title = {Guidelines for the {Evolving} {Role} of {Generative} {AI} in {Introductory} {Programming} {Based} on {Emerging} {Practice}},
    isbn = {979-8-4007-0600-4},
    url = {https://dl.acm.org/doi/10.1145/3649217.3653602},
    doi = {10.1145/3649217.3653602},
    urldate = {2025-09-03},
    booktitle = {Proceedings of the 2024 on {Innovation} and {Technology} in {Computer} {Science} {Education} {V}. 1},
    publisher = {Association for Computing Machinery},
    author = {Mahon, Joyce and Mac Namee, Brian and Becker, Brett A.},
    year = {2024},
    pages = {10--16},
}

@article{shoufan_exploring_2023,
    title = {Exploring {Students}’ {Perceptions} of {ChatGPT}: {Thematic} {Analysis} and {Follow}-{Up} {Survey}},
    volume = {11},
    issn = {2169-3536},
    shorttitle = {Exploring {Students}’ {Perceptions} of {ChatGPT}},
    url = {https://ieeexplore.ieee.org/document/10105236},
    doi = {10.1109/ACCESS.2023.3268224},
    urldate = {2025-09-04},
    journal = {IEEE Access},
    author = {Shoufan, Abdulhadi},
    year = {2023},
    pages = {38805--38818},
}

@inproceedings{margulieux_self-regulation_2024,
    address = {New York, NY, USA},
    series = {{ITiCSE} 2024},
    title = {Self-{Regulation}, {Self}-{Efficacy}, and {Fear} of {Failure} {Interactions} with {How} {Novices} {Use} {LLMs} to {Solve} {Programming} {Problems}},
    isbn = {979-8-4007-0600-4},
    url = {https://dl.acm.org/doi/10.1145/3649217.3653621},
    doi = {10.1145/3649217.3653621},
    urldate = {2025-09-02},
    booktitle = {Proceedings of the 2024 on {Innovation} and {Technology} in {Computer} {Science} {Education} {V}. 1},
    publisher = {Association for Computing Machinery},
    author = {Margulieux, Lauren E. and Prather, James and Reeves, Brent N. and Becker, Brett A. and Cetin Uzun, Gozde and Loksa, Dastyni and Leinonen, Juho and Denny, Paul},
    year = {2024},
    pages = {276--282},
}

@inproceedings{tankelevitch_metacognitive_2024,
    address = {New York, NY, USA},
    series = {{CHI} '24},
    title = {The {Metacognitive} {Demands} and {Opportunities} of {Generative} {AI}},
    isbn = {979-8-4007-0330-0},
    url = {https://dl.acm.org/doi/10.1145/3613904.3642902},
    doi = {10.1145/3613904.3642902},
    urldate = {2025-09-03},
    booktitle = {Proceedings of the 2024 {CHI} {Conference} on {Human} {Factors} in {Computing} {Systems}},
    publisher = {Association for Computing Machinery},
    author = {Tankelevitch, Lev and Kewenig, Viktor and Simkute, Auste and Scott, Ava Elizabeth and Sarkar, Advait and Sellen, Abigail and Rintel, Sean},
    year = {2024},
    pages = {1--24},
}

@misc{yuan_generative_2024,
    title = {Generative {AI} as a {Tool} for {Enhancing} {Reflective} {Learning} in {Students}},
    url = {https://arxiv.org/abs/2412.02603v1},
    doi = {https://doi.org/10.48550/arXiv.2412.02603},
    language = {en},
    urldate = {2025-06-19},
    author = {Yuan, Bo and Hu, Jiazi},
    month = nov,
    year = {2024},
}

@inproceedings{denny_explaining_2024,
    address = {New York, NY, USA},
    series = {{ITiCSE} 2024},
    title = {Explaining {Code} with a {Purpose}: {An} {Integrated} {Approach} for {Developing} {Code} {Comprehension} and {Prompting} {Skills}},
    isbn = {979-8-4007-0600-4},
    shorttitle = {Explaining {Code} with a {Purpose}},
    url = {https://dl.acm.org/doi/10.1145/3649217.3653587},
    doi = {10.1145/3649217.3653587},
    urldate = {2025-09-03},
    booktitle = {Proceedings of the 2024 on {Innovation} and {Technology} in {Computer} {Science} {Education} {V}. 1},
    publisher = {Association for Computing Machinery},
    author = {Denny, Paul and Smith, David H. and Fowler, Max and Prather, James and Becker, Brett A. and Leinonen, Juho},
    year = {2024},
    pages = {283--289},
}

@inproceedings{denny_prompt_2024,
    address = {New York, NY, USA},
    series = {{SIGCSE} 2024},
    title = {Prompt {Problems}: {A} {New} {Programming} {Exercise} for the {Generative} {AI} {Era}},
    isbn = {979-8-4007-0423-9},
    shorttitle = {Prompt {Problems}},
    url = {https://dl.acm.org/doi/10.1145/3626252.3630909},
    doi = {10.1145/3626252.3630909},
    urldate = {2025-09-02},
    booktitle = {Proceedings of the 55th {ACM} {Technical} {Symposium} on {Computer} {Science} {Education} {V}. 1},
    publisher = {Association for Computing Machinery},
    author = {Denny, Paul and Leinonen, Juho and Prather, James and Luxton-Reilly, Andrew and Amarouche, Thezyrie and Becker, Brett A. and Reeves, Brent N.},
    year = {2024},
    pages = {296--302},
}

@article{akmam_integration_2024,
    title = {Integration of cognitive conflict in generative learning model to enhancing students’ creative thinking skills},
    volume = {20},
    issn = {1305-8215, 1305-8223},
    url = {https://www.ejmste.com/article/integration-of-cognitive-conflict-in-generative-learning-model-to-enhancing-students-creative-15026},
    doi = {10.29333/ejmste/15026},
    language = {english},
    number = {9},
    urldate = {2025-06-12},
    journal = {Eurasia Journal of Mathematics, Science and Technology Education},
    author = {Akmam, Akmam and Afrizon, Renol and Koto, Irwan and Setiawan, David and Hidayat, Rahmat and Novitra, Fuja},
    month = sep,
    year = {2024},
    note = {Publisher: Modestum},
    pages = {em2504},
}

@inproceedings{prasad_self-regulated_2024,
    address = {New York, NY, USA},
    series = {{SIGCSE} 2024},
    title = {A {Self}-{Regulated} {Learning} {Framework} using {Generative} {AI} and its {Application} in {CS} {Educational} {Intervention} {Design}},
    isbn = {979-8-4007-0423-9},
    url = {https://dl.acm.org/doi/10.1145/3626252.3630828},
    doi = {10.1145/3626252.3630828},
    urldate = {2025-09-03},
    booktitle = {Proceedings of the 55th {ACM} {Technical} {Symposium} on {Computer} {Science} {Education} {V}. 1},
    publisher = {Association for Computing Machinery},
    author = {Prasad, Prajish and Sane, Aamod},
    year = {2024},
    pages = {1070--1076},
}

@inproceedings{kazemitabaar_improving_2024,
    address = {New York, NY, USA},
    series = {{UIST} '24},
    title = {Improving {Steering} and {Verification} in {AI}-{Assisted} {Data} {Analysis} with {Interactive} {Task} {Decomposition}},
    isbn = {979-8-4007-0628-8},
    url = {https://dl.acm.org/doi/10.1145/3654777.3676345},
    doi = {10.1145/3654777.3676345},
    urldate = {2025-09-02},
    booktitle = {Proceedings of the 37th {Annual} {ACM} {Symposium} on {User} {Interface} {Software} and {Technology}},
    publisher = {Association for Computing Machinery},
    author = {Kazemitabaar, Majeed and Williams, Jack and Drosos, Ian and Grossman, Tovi and Henley, Austin Zachary and Negreanu, Carina and Sarkar, Advait},
    year = {2024},
    pages = {1--19},
}

@article{braun_using_2006,
    title = {Using thematic analysis in psychology},
    volume = {3},
    issn = {1478-0887},
    url = {https://www.tandfonline.com/doi/abs/10.1191/1478088706qp063oa},
    doi = {10.1191/1478088706qp063oa},
    number = {2},
    urldate = {2024-03-14},
    journal = {Qualitative Research in Psychology},
    author = {Braun, Virginia and Clarke, Victoria},
    month = jan,
    year = {2006},
    note = {Publisher: Routledge
\_eprint: https://www.tandfonline.com/doi/pdf/10.1191/1478088706qp063oa},
    pages = {77--101},
}

@article{joshi_likert_2015,
    title = {Likert {Scale}: {Explored} and {Explained}},
    volume = {7},
    issn = {22310843},
    shorttitle = {Likert {Scale}},
    url = {https://journalcjast.com/index.php/CJAST/article/view/381},
    doi = {10.9734/BJAST/2015/14975},
    language = {en},
    number = {4},
    urldate = {2025-09-05},
    journal = {British Journal of Applied Science \& Technology},
    author = {Joshi, Ankur and Kale, Saket and Chandel, Satish and Pal, D.},
    month = jan,
    year = {2015},
    pages = {396--403},
}

@book{paul_thinkers_2019,
    address = {London, UK},
    title = {The {Thinker}'s {Guide} to {Socratic} {Questioning}},
    isbn = {978-1-5381-3381-1},
    language = {en},
    publisher = {Bloomsbury Publishing PLC},
    author = {Paul, Richard and Elder, Linda},
    month = jun,
    year = {2019},
    note = {Google-Books-ID: ADWbDwAAQBAJ},
}

@inproceedings{lee_impact_2025,
    address = {New York, NY, USA},
    series = {{CHI} '25},
    title = {The {Impact} of {Generative} {AI} on {Critical} {Thinking}: {Self}-{Reported} {Reductions} in {Cognitive} {Effort} and {Confidence} {Effects} {From} a {Survey} of {Knowledge} {Workers}},
    isbn = {979-8-4007-1394-1},
    shorttitle = {The {Impact} of {Generative} {AI} on {Critical} {Thinking}},
    url = {https://dl.acm.org/doi/10.1145/3706598.3713778},
    doi = {10.1145/3706598.3713778},
    urldate = {2025-11-06},
    booktitle = {Proceedings of the 2025 {CHI} {Conference} on {Human} {Factors} in {Computing} {Systems}},
    publisher = {Association for Computing Machinery},
    author = {Lee, Hao-Ping (Hank) and Sarkar, Advait and Tankelevitch, Lev and Drosos, Ian and Rintel, Sean and Banks, Richard and Wilson, Nicholas},
    year = {2025},
    pages = {1--22},
}

@inproceedings{prather_widening_2024,
    address = {New York, NY, USA},
    series = {{ICER} '24},
    title = {The {Widening} {Gap}: {The} {Benefits} and {Harms} of {Generative} {AI} for {Novice} {Programmers}},
    volume = {1},
    isbn = {979-8-4007-0475-8},
    shorttitle = {The {Widening} {Gap}},
    url = {https://dl.acm.org/doi/10.1145/3632620.3671116},
    doi = {10.1145/3632620.3671116},
    urldate = {2025-11-06},
    booktitle = {Proceedings of the 2024 {ACM} {Conference} on {International} {Computing} {Education} {Research} - {Volume} 1},
    publisher = {Association for Computing Machinery},
    author = {Prather, James and Reeves, Brent N and Leinonen, Juho and MacNeil, Stephen and Randrianasolo, Arisoa S and Becker, Brett A. and Kimmel, Bailey and Wright, Jared and Briggs, Ben},
    year = {2024},
    pages = {469--486},
}

@inproceedings{amoozadeh_student-ai_2024,
    address = {New York, NY, USA},
    series = {Koli {Calling} '24},
    title = {Student-{AI} {Interaction}: {A} {Case} {Study} of {CS1} students},
    isbn = {979-8-4007-1038-4},
    shorttitle = {Student-{AI} {Interaction}},
    url = {https://dl.acm.org/doi/10.1145/3699538.3699567},
    doi = {10.1145/3699538.3699567},
    urldate = {2025-11-06},
    booktitle = {Proceedings of the 24th {Koli} {Calling} {International} {Conference} on {Computing} {Education} {Research}},
    publisher = {Association for Computing Machinery},
    author = {Amoozadeh, Matin and Nam, Daye and Prol, Daniel and Alfageeh, Ali and Prather, James and Hilton, Michael and Srinivasa Ragavan, Sruti and Alipour, Amin},
    year = {2024},
    pages = {1--13},
}

@inproceedings{lau_ban_2023,
    address = {New York, NY, USA},
    series = {{ICER} '23},
    title = {From "{Ban} {It} {Till} {We} {Understand} {It}" to "{Resistance} is {Futile}": {How} {University} {Programming} {Instructors} {Plan} to {Adapt} as {More} {Students} {Use} {AI} {Code} {Generation} and {Explanation} {Tools} such as {ChatGPT} and {GitHub} {Copilot}},
    volume = {1},
    isbn = {978-1-4503-9976-0},
    shorttitle = {From "{Ban} {It} {Till} {We} {Understand} {It}" to "{Resistance} is {Futile}"},
    url = {https://dl.acm.org/doi/10.1145/3568813.3600138},
    doi = {10.1145/3568813.3600138},
    language = {en-US},
    urldate = {2025-11-06},
    booktitle = {Proceedings of the 2023 {ACM} {Conference} on {International} {Computing} {Education} {Research} - {Volume} 1},
    publisher = {Association for Computing Machinery},
    author = {Lau, Sam and Guo, Philip},
    year = {2023},
    pages = {106--121},
}

@article{jost_impact_2024,
    title = {The {Impact} of {Large} {Language} {Models} on {Programming} {Education} and {Student} {Learning} {Outcomes}},
    volume = {14},
    copyright = {http://creativecommons.org/licenses/by/3.0/},
    issn = {2076-3417},
    url = {https://www.mdpi.com/2076-3417/14/10/4115},
    doi = {10.3390/app14104115},
    language = {en},
    number = {10},
    pages = {4115},
    urldate = {2025-11-06},
    journal = {Applied Sciences},
    author = {Jošt, Gregor and Taneski, Viktor and Karakatič, Sašo and Jošt, Gregor and Taneski, Viktor and Karakatič, Sašo},
    month = may,
    year = {2024},
    note = {Company: Multidisciplinary Digital Publishing Institute
Distributor: Multidisciplinary Digital Publishing Institute
Institution: Multidisciplinary Digital Publishing Institute
Label: Multidisciplinary Digital Publishing Institute
Publisher: publisher},
}

@inproceedings{ghimire_generative_2024,
    address = {Washington, DC, USA},
    title = {Generative {AI} in {Education}: {A} {Study} of {Educators}' {Awareness}, {Sentiments}, and {Influencing} {Factors}},
    shorttitle = {Generative {AI} in {Education}},
    url = {https://ieeexplore.ieee.org/abstract/document/10892891},
    doi = {10.1109/FIE61694.2024.10892891},
    language = {en-US},
    urldate = {2025-11-06},
    booktitle = {2024 {IEEE} {Frontiers} in {Education} {Conference} ({FIE})},
    publisher = {IEEE},
    author = {Ghimire, Aashish and Pather, James and Edwards, John},
    month = oct,
    year = {2024},
    note = {ISSN: 2377-634X},
    pages = {1--9},
}

@article{nathaniel_investigating_2025,
    title = {Investigating the impact of generative {AI} integration on the sustenance of higher-order thinking skills and understanding of programming logic},
    volume = {9},
    issn = {2666920X},
    url = {https://linkinghub.elsevier.com/retrieve/pii/S2666920X25001006},
    doi = {10.1016/j.caeai.2025.100460},
    language = {en},
    urldate = {2025-11-06},
    journal = {Computers and Education: Artificial Intelligence},
    author = {Nathaniel, Jemimah and Oyelere, Solomon Sunday and Suhonen, Jarkko and Tedre, Matti},
    month = dec,
    year = {2025},
    pages = {100460},
}

@article{kong_pedagogical_2024,
    title = {A pedagogical design for self-regulated learning in academic writing using text-based generative artificial intelligence tools: 6-{P} pedagogy of plan, prompt, preview, produce, peer-review, portfolio-tracking},
    volume = {19},
    issn = {1793-7078},
    shorttitle = {A pedagogical design for self-regulated learning in academic writing using text-based generative artificial intelligence tools},
    url = {https://rptel.apsce.net/index.php/RPTEL/article/view/2024-19030},
    doi = {10.58459/rptel.2024.19030},
    language = {en},
    urldate = {2025-11-17},
    journal = {Research and Practice in Technology Enhanced Learning},
    author = {Kong, Siu-Cheung and Lee, John Chi-Kin and Tsang, Olson},
    month = jan,
    year = {2024},
    pages = {030--030},
}

@misc{zhou_teachers_2024,
    title = {"{The} teachers are confused as well": {A} {Multiple}-{Stakeholder} {Ethics} {Discussion} on {Large} {Language} {Models} in {Computing} {Education}},
    shorttitle = {"{The} teachers are confused as well"},
    url = {http://arxiv.org/abs/2401.12453},
    doi = {10.48550/arXiv.2401.12453},
    urldate = {2026-01-26},
    publisher = {arXiv},
    author = {Zhou, Kyrie Zhixuan and Kilhoffer, Zachary and Sanfilippo, Madelyn Rose and Underwood, Ted and Gumusel, Ece and Wei, Mengyi and Choudhry, Abhinav and Xiong, Jinjun},
    month = jan,
    year = {2024},
    note = {arXiv:2401.12453 [cs]},
}

@article{li_large_2024,
    title = {Large language models and medical education: a paradigm shift in educator roles},
    volume = {11},
    issn = {2196-7091},
    shorttitle = {Large language models and medical education},
    url = {https://doi.org/10.1186/s40561-024-00313-w},
    doi = {10.1186/s40561-024-00313-w},
    language = {en},
    number = {1},
    urldate = {2026-01-26},
    journal = {Smart Learning Environments},
    author = {Li, Zhui and Li, Fenghe and Fu, Qining and Wang, Xuehu and Liu, Hong and Zhao, Yu and Ren, Wei},
    month = jun,
    year = {2024},
    pages = {26},
}

\appendix

\section{Ethical concerns - Irresponsible contents}\label{A:Ethical}
%7个学生认为在用LLM学习时没遇到不负责任的问题，其他学生认为用LLM学习时主要面临幻觉问题。5个学生（P2，P5，P7，P9，P11）认为LLM生成的代码和题解会出现幻觉，3个学生（P8，P13，P15）认为生成文章中可能出现"不存在的文献"。没有学生认为自己受到幻觉影响，他们都表示LLMs存在幻觉众所周知，平时就会注意。而当问及LLMs添加护栏而只能引导学生时，4个学生（P2，P7，P11，P16）认为即便LLM不提供答案只反问也可能出现幻觉，因为LLMs只要回应就会有幻觉，只是这种情况下幻觉的可能性比直接生成答案来的少。
Seven students reported no issues of irresponsibility when using LLMs for learning, while the others said the main challenge was hallucination. Five students (P2/5/7/9/11) noted hallucinations in LLM-generated code and quiz solutions, and three (P8/13/15) reported "nonexistent references" in generated writing. No student felt personally misled by hallucinations because hallucinations in LLMs are well known, and they stay alert. When asked about adding guardrails so that LLMs only guide students, four students (P2/7/11/16) argued that hallucinations can still occur even if the LLM withholds answers and only asks counter-questions, because hallucinations can appear whenever the model responds, though they considered the likelihood lower than when it directly generates answers.

\section{Changes in relationships and roles}\label{A:Relation}
%既往研究表明，LLM的问世深刻重塑了师生关系：凭借即时响应的优势，LLMs正逐渐取代传统辅导角色，成为学生答疑解惑的首选 \cite{chowdhury_autotutor_2024, hou_effects_2024}。因此，教师的角色从单纯的知识传授者转变为情感支持者 \cite{chan_will_2024, li_large_2024}、AI工具使用的审核和引导者 \cite{chan_will_2024, ghimire_generative_2024, zhou_teachers_2024}。这种转变不仅填补了教师无法随时在线的空白，更使教师得以将重复性教学工作“外包”给LLM，从而专注于深层指导 \cite{kumar_guiding_2024, mogavi_chatgpt_2024}。在此基础上，本研究进一步明确了不同场景与意图下师生和LLMs的关系图景。

%在写作和有明确答案的意图下 - "(Programming) quizzes-solving"，"(Programming) quizzes-correction"和"Info-query \& summary" - 除少数学生（P5/8/9）外，绝大多数参与者倾向于优先咨询LLMs。然而，这并不意味着LLMs完全取代了教师；当LLMs无法解决问题或存在潜在错误时，8名学生表示会立即转向教师寻求帮助。这种“LLMs优先，教师兜底”的模式也得到了所有受访教师的认可：这种分流机制有效减轻了他们的负担，使其能集中精力解决LLM难以处理的教学难点。

%在基于项目的学习等开放式任务中，教师与LLMs被认为发挥着重要的互补作用。部分学生（P3/5/10/14）认为教师在构思的可行性上优于LLMs，而另一部分（P6/12/16）则认为LLMs更有助于发散思维。T4进一步期望学生先利用LLM进行初步构思，再由教师提供反馈。而且，在涉及高度专业经验的任务（如制定研究方案）中，大多数师生（6名学生，5名教师）一致认为教师的指导作用不可替代。

%最后，LLM无法取代教师在教学设计层面的职能。T1指出，学生通过LLMs的学习往往是被动的，依赖于教师的安排，而LLM本身无法进行系统化的教学安排与主动授课（5名学生；T1/2/3）。T2和T4强调，教师的独特性体现在教育设计的本质上——即优质的课程设计、教学策略与评估方式只能由教师主导。不过，T2也表达了对未来的担忧：随着这些创新教学模式被LLMs习得，教师在这一领域的不可替代性未来或许也会面临挑战。

Previous research indicates that LLMs have profoundly reshaped the instructor-student relationship: leveraging their advantage of instant response, LLMs are gradually replacing traditional tutoring roles, becoming the primary choice for answering student queries \cite{chowdhury_autotutor_2024, hou_effects_2024}. Consequently, the instructor's role has shifted from a knowledge transmitter to an emotional supporter \cite{chan_will_2024, li_large_2024} and an evaluator and guide for AI tool usage \cite{chan_will_2024, ghimire_generative_2024, zhou_teachers_2024}. This shift not only fills the gap of instructor unavailability but also allows instructors to "outsource" repetitive teaching tasks to LLMs, thereby enabling a focus on deeper guidance \cite{kumar_guiding_2024, mogavi_chatgpt_2024}. Building on this foundation, our study further clarifies the relational landscape between students, instructors, and LLMs across specific scenarios and intentions.

\textbf{The vast majority of participants (except for P5/8/9) tended to consult LLMs first in writing and intents with definitive answers} - "(Programming) quizzes-solving", "(Programming) quizzes-correction", and "Info-query \& summary". However, this does not imply that LLMs have replaced instructors; when LLMs fail to solve a problem or provide potentially erroneous answers, eight students stated they would immediately turn to instructors for help. This "LLMs-first, instructor-fallback" pattern was acknowledged by all interviewed instructors: this mechanism effectively reduces their burden, allowing them to concentrate on difficulties that LLMs struggle to address.

\textbf{In open-ended tasks such as project-based learning, instructors and LLMs are perceived to play important complementary roles}. Some students (P3/5/10/14) believed instructors were superior to LLMs regarding the feasibility of ideas, while others (P6/12/16) felt LLMs were more helpful for divergent thinking. T4 further expected that students should first use LLMs for preliminary ideation, followed by instructor feedback. Moreover, in tasks involving high-level professional experience (e.g., formulating research proposals), the majority of students and instructors (6 students; 5 instructors) agreed that the instructor's advice is irreplaceable.

Finally, \textbf{LLMs cannot replace the instructor's function in instructional design}. T1 pointed out that student learning via LLMs is often passive, relying on instructor arrangements, as LLMs themselves cannot conduct systematic pedagogical planning or active instruction (5 students; T1/2/3). T2 and T4 emphasized that the instructor's uniqueness lies in the essence of educational design - high-quality curriculum design, pedagogical strategies, and assessment methods can only be led by instructors. However, T2 also expressed concern about the future: as these innovative teaching models are learned by LLMs, the instructor's role in this domain might also face challenges.

\end{document}